\documentclass[twocolumn,prb,amsmath,amssymb]{revtex4}
\usepackage{graphicx,bm,float,color,hyperref}
\hypersetup{colorlinks=true linkcolor={blue} }
\def\<{{\langle}}
\def\>{{\rangle}}

\newcommand{\al}{\alpha}
\newcommand{\be}{\beta}
\newcommand{\de}{\delta}
\newcommand{\D}{\Delta}
\newcommand{\e}{\epsilon}
\newcommand{\bare}{\bar{\e}}

\newcommand{\lm}{\lambda}

\newcommand{\w}{\omega}
\newcommand{\W}{\Omega}

\newcommand{\s}{\sigma}

\newcommand{\del}{\nabla}

\newcommand{\p}{\partial}

\newcommand{\bsdel}{\boldsymbol{\nabla}}
\newcommand{\va}{\vec{a}}
\newcommand{\vpsi}{\vec{\psi}}

\newcommand{\vJ}{\vec{J}}

\newcommand{\mbd}{\mathbf{d}}
\newcommand{\mbh}{\mathbf{h}}

\newcommand{\mbk}{\mathbf{k}}
\newcommand{\mbK}{\mathbf{K}}

\newcommand{\ksn}{{\mbk s n}}

\newcommand{\mbq}{\mathbf{q}}

\newcommand{\mbr}{\mathbf{r}}

\newcommand{\mbs}{\mathbf{s}}

\newcommand{\mbv}{\mathbf{v}}

\newcommand{\bsphi}[1]{\boldsymbol{\phi}}
\newcommand{\mbf}[1]{\mathbf{#1}}

\newcommand{\hsig}{\hat{\sigma}}
\newcommand{\bfhat}[1]{\hat{\mathbf{{#1}}}}

\newcommand{\mcal}[1]{\mathcal{#1}}

\newcommand{\til}[1]{\tilde{#1}}
\newcommand{\pfrac}[2]{\left(\frac{#1}{#2}\right)}

\newcommand{\hatA}{\hat{A}}
\newcommand{\hatG}{\hat{G}}
\newcommand{\hatU}{\hat{U}}
\newcommand{\hath}{\hat{h}}

\newcommand{\pauli}{\hat{\boldsymbol{\sigma}}}

\newcommand{\nn}{\nonumber\\}

\newcommand{\ben}{\begin{equation}}
\newcommand{\een}{\end{equation}}

\newcommand{\fig}[1]{Fig.~\ref{#1} }
\newcommand{\mbhk}{\mbh_\mbk}
\newcommand{\dkk}{{d\mathbf{k}\over(2\pi)^{2}}}
\begin{document}
\title{Quasiparticle Berry curvature and Chern numbers in spin-orbit coupled bosonic Mott insulators}
\author{Clement H. Wong and R.A. Duine}
\affiliation{Institute for Theoretical Physics, Utrecht University, Leuvenlaan 4, 3584 CE Utrecht, The Netherlands}
\date{\today}
\begin{abstract}
We study the ground state topology and quasiparticle properties in bosonic Mott insulators with two dimensional spin-orbit couplings in cold atomic optical lattices.  We show that the many-body Chern and spin-Chern number can be expressed as an integral of the quasihole Berry curvatures over the Brillouin zone.  Using a strong coupling perturbation theory, for an experimentally feasible spin-orbit coupling, we compute the Berry curvature and the spin Chern number and find that these quantities can be generated purely by interactions.  We also compute the quasiparticle dispersions, spectral weights, and the quasimomentum space distribution of particle and spin density, which can be accessed in cold atom experiments and used to deduce the Berry curvature and Chern numbers.
\end{abstract}

\maketitle
%\tableofcontents
\section{introduction}
Spin-orbit (SO) coupling is responsible for many Hall and quantum Hall type phenomena and associated topological phases in solid state systems without magnetic fields, such as the anomalous Hall effect in ferromagnetic metals, the spin Hall effect in semiconductors, and the quantum spin Hall effect in topological insulators.\cite{hasanRMP10}  Recently, there has been much experimental effort to engineer spin-orbit couplings in cold atomic systems, with the aim of achieving similar topological phases, such as the quantum anomalous Hall insulator,\cite{liuPRA10,liuCM13}  and the atomic topological insulators and superfluids.\cite{goldmanPRL10, sauPRB11}

While topological insulators and related phases are well understood in the weakly interacting limit,\cite{hasanRMP10} recent work has turned toward the strongly interacting regime,\cite{wangPRL10to,yoshidaPRB12} which is less well understood and may be relevant for some strongly correlated materials such as the transition metal oxides.  An outgrowth of this field is the study of bosonic topological insulators and related quantum phases,\cite{chenPRB11,senthilPRL13}which, unlike fermions, necessarily require interactions.  Ultracold atomic systems are the natural candidates for creating these quantum phases because of the ability of experimentalists to trap bosons in clean optical lattices and to control microscopic parameters such as hopping and interaction strength.\cite{greinerNAT02} 

So far, only one-dimensional (abelian) SO couplings equivalent to a combination of Rashba and Dresselhaus SO coupling with equal magnitude has been achieved in cold atom experiments.\cite{linNAT11}    Experimental methods to achieve two-dimensional SO coupling  (i.e. Rashba) have been proposed,\cite{liuCM13}but topologically nontrivial phases generally require a three dimensional SO coupling which is much more difficult to realize without introducing a sublattice degree of freedom. On the other hand, as we demonstrate in this paper, interactions in the ferromagnetic Mott insulating regime can generate a SO coupling in the quasiparticle  Hamiltonian that is independent of the SO coupling in the hopping Hamiltonian, providing another experimental knob to engineer SO coupling.   
 
With these motivations, building on our previous work in Ref.~[\onlinecite{wongPRL13}], we study and compute the ground states properties of SO coupled bosonic Mott insulators, including the quasiparticle dispersions and spectral weights, the quasimomentum space distributions of particle and spin density, the interacting SO texture, Berry curvature, and Chern numbers.  We present a strong coupling perturbation theory to compute the single particle propagator in the Mott insulator, from which the aforementioned ground-state properties will follow.  We will furthermore show that the interaction generated Berry curvature can lead to a topological phase characterized by a spin Chern number.

Our results can be measured by various experimental techniques. The quasimomentum space distributions of particle and spin density can be measured using phase contrast imaging followed by time of flight measurements.\cite{sauPRB11} The interacting SO texture and the corresponding  Berry curvature can be inferred from the these measurements, once the many body occupation function on the quasiparticle SO bands which determine the local quasimomentum space polarization is known, which, as we will show, can be computed in our formalism.  

For the cases considered in this paper, the Chern numbers will depend only on the sign of the SO texture at four high symmetry points in the Brillouin zone.  In fact, we will show that at these points, the SO texture is proportional to the SO energy splitting in the quasiparticle dispersions [cf.~section\ref{fkz}], which can be measured for example, with Bragg spectroscopy, providing a simple way of experimentally determining the spin Chern number.

The quasiparticle Berry curvature causes an anomalous velocity in the wave packet dynamics that can be measured as deviations in Bloch oscillations, which, in the absence of Berry curvature, are real space oscillations of quasiparticles wave packets in the direction of the external forces, stemming from the usual band velocity in quasimomentum space.  Recently, there have been several experimental proposals for measuring these deviations for the noninteracting Berry curvature,\cite{pricePRA12,liuCM13} and we expect similar methods can be applied to measure the interaction-generated ones that we study as well.

This paper is organized as follows. In section \ref{hopping}, we introduce the SO coupled Bose Hubbard model and review the hopping Hamiltonian we use for our calculations.  In section \ref{Gfcn}, we review the properties of the single particle propagator in the Mott insulator.  In section \ref{berrychern}, we show that the interacting Chern number can be expressed in terms of the Berry curvature of the quaisparticle Hamiltonian.   In section \ref{pert}, we compute the propagator using the strong coupling perturbation theory.  In section \ref{results} we compute the aforementioned observables for on-site parameters given in section \ref{without}, and for a ferromagnetic Mott insulating state, we compute the phase diagram determined by the spin Chern number.

\section{Bose Hubbard model with spin-orbit coupling }
\label{hopping}
We consider a two dimensional optical lattice of bosons with two spin components, generic spin-dependent hopping amplitudes with onsite repulsive interactions, deep in the Mott-insulating phase at a commensurate filling.
 We will describe the system by a  pseudo-spin $1/2$  Bose Hubbard  model and write the Hamiltonian as
\[H=H_0+V\,,\]  
where
\begin{align*}
H_0&={1\over 2} \sum_{i,\al\beta} U_{\al\beta}a^\dag_{i\al}a^\dag_{i\beta} a_{i\beta}a_{i\al}-\mu_\al a^\dag_{i\al}a_{i\al}\nn
&=\sum_{i\al} \left[{U_{\al\al}\over 2}n_{i\al}(n_{i\al}-1)-\mu_\al n_{i\al}\right]+U_{+-}n_{i+}n_{i-} \,,
\label{H0}
\end{align*}
and
\[
V=\sum_{i,j} \va_{i}^\dag(\hat{\e}\de_{ij}+\hat{t}_{ij}) \va_{j}\,,
\]
where  $i,j$ denotes the position of atoms in the optical lattice, $\al,\beta=\pm$ are spin indices, hats denote matrices in spin space, $\vec{a}_i$ and $n_{i\al}=a^\dag_{i\al}{a}_{i\al}$ are two-component spinor field operators and occupation number operators in the real space Wannier basis, respectively, and we will keep only the lowest Bloch band.  Also, $\hat{\e}$ are the onsite energies, and $\hat{t}_{ij}$ are the matrices of hopping amplitudes. 

We will develop our theory for a generic tight binding Hamiltonian with SO or pseudo-SO coupling, and illustrate the results with the SO coupling which was recently proposed for a cold atom system which may have some experimental advantages.\cite{liuCM13} This SO coupling is generated in a square optical lattice and is formally identical to one component of the Bernevig--Hughes--Zhang  Hamiltonian which describes Mercury Telluride quantum wells, the well known solid state topological insulator.\cite{bernevigSCI06}  Of course, the microscopic meaning of the parameters are different; specifically, the spin in our model refers to the atomic hyperfine states.

The quasimomentum space hopping Hamiltonian is given by 
\[
V=\int_{\rm BZ} {d\mbk\over(2\pi)^2}\,
 \va_{\mbk}^\dag \hat{h}_\mbk\va_{\mbk}\,,\quad 
\hat{h}_\mbk=h_\mbk+\mbf{h}_\mbk\cdot\hat{\bm\s}\,, 
\]
where the quasi-momentum $\mbk$ is in units of inverse lattice spacing, the integral is over the Brillouin zone (BZ),   $\vec{a}_\mbk$ denote second quantized operators in quasi--momentum space, given by $\va_\mbk=\sum_i \va_ie^{i\mbk\cdot\mbr_i}$,\footnote{We impose real space periodic boundary conditions and take the limit of infinite lattices sites $N_s\to\infty$, so that 
\[\de_{ij}=\int \dkk e^{i\mbk\cdot(\mbr_i-\mbr_j)}\,,\quad (2\pi)^2\de(\mbk-\mbk')=\sum_i^\infty e^{i(\mbk-\mbk')\cdot\mbr_i}\]} where $\{\mbr_i\}$ are the bravais lattice vectors.   where $\hat{\bm{\s}}$ is the vector of Pauli matrices, the non-interacting spin-orbit field is given by $\mbf{h}_\mbk$,  and the dispersions of the non-interacting particles are given by $\e^{(1)}_{\mbk\al}=h_\mbk+\al|\mbf{h}_\mbk|$.  
\begin{figure*}[t]
\includegraphics[width=.28\linewidth]{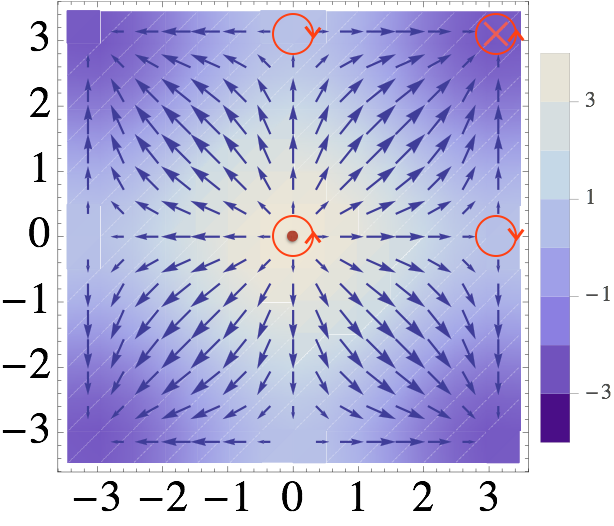}
\qquad
\begin{tabular}[b]{ccccccc}
\hline\hline
$\mbk_i$&$c(\mbk_i)$&$p(\mbk_i)c(\mbk_i)$ & $0\leq\D<8B$&$-8B<\D\leq0$&$\D<-8B$& $\D>8B$ \\
\hline
$(0,0)$&+&sgn$(\D/2+4B)$ & $+$& $+$& $-$& $+$ \\
$(\pi,\pi)$&+&sgn $(\D/2-4B)$ &$-$& $-$& $-$& $+$ \\
$(0,\pi),(\pi,0)$&$-$&$-$sgn$(\D/2)$ & $-$& $+$& $+$& $+$ \\
&&$\sum_ip(\mbk_i)c(\mbk_i)/2$&$-1$& $1$& $0$& $0$ \\
\hline\hline
\end{tabular}
\caption{(Color Online) Left: Spin-orbit texture of the hopping Hamiltonian, $\mbh_\mbk/t$ for $A=B=t$, $\D=0$ where  the vector fields and and density plot denote in-plane components $(h_x,h_y)$ and z-component $h_z/U$, respectively.  Oriented contours in red centered around vortices at indicated the chirality and polarity.  Right:  Computation of the sum of topological charges of quasimomentum space vortices as a function of $\Delta,B$ in the noninteracting BHZ model.}
\label{topcharge}
\end{figure*}

In our model, we will consider only nearest neighbor hopping.  The hopping matrices along the unit vectors of the $x,y$ axes $\{\mbf{e}_\de\}=\{\pm\bfhat{x},\pm\bfhat{y}\}$ with the average onsite energies set to zero are given by
\begin{widetext}
\[
 \hat{\e}=\left(
\begin{array}{cc}
 \D/2 & 0\\
 0& -\D/2
\end{array}\right)\,,\quad
 \hat{t}_{\pm x}=\left(
\begin{array}{cc}
D+B & \pm i A \\
 \pm i A & D-B
\end{array}
\right)\,,\quad
\hat{t}_{\pm y}=\left(
\begin{array}{cc}
D+B & \pm  A \\
 \mp  A &D-B
\end{array}
\right)\,.
\]

It is convenient to parametrize the matrices of hopping amplitudes of by $\hat{t}_\de={t}_\de+\mbf{t}_\de\cdot\hat{\bm{\s}}$, then $
\hath_\mbk=\hat{\e}+\sum_\de \hat{t}_\de e^{-i\mbk\cdot\mbf{e}_\de}$, or, in components 
$h_\mbk=\hat{\e}+\sum_\de {t}_\de e^{-i\mbk\cdot\mbf{e}_\de}\,,
 \mbh_\mbk=\sum_\de\mbf{t}_\de e^{-i\mbk\cdot\mbf{e}_\de}\,,$
from which follows
\ben
h_\mbk=2D(\cos{k_x}+ \cos{k_y})\,,\quad
(h_x,h_y)=2A(\sin{k_x},\sin{k_y})\,, \quad
h_z={\D\over2}+2B(\cos{k_x}+ \cos{k_y})\,.
\label{hBHZ}
\een

Next, we give a brief review of this model in the absence of interactions, which contains many features that will remain in the Mott insulator phase.  As evident from the $x,y$ hopping matrices and $h_{x,y}(\mbk)$, this model breaks inversion symmetry (defined by $\mathbf{k}\to-\mathbf{k}$), 
except at special points $\mbk_i=\{(0,0),(\pi,\pi),(\pi,0),(0,\pi)\}$, which correspond to locations of  vortices ($\{(\pi,0),(0,\pi)\}$) and anti-vortices ($\{(0,0),(\pi,\pi)\}$) in the SO field $\mbh_\mbk$, plotted for $A=B$ in Fig.~\ref{topcharge}(left). If one defines an artificial time reversal symmetry operator,\footnote{The actual time reversal operator is given by $\mcal{T}=e^{-i\pi S_y/\hbar}K$, with $S_y$ refering to the spin of the total hyperfine multiplet, $K$ being complex conjugation.} 
$\mcal{T}=i\hat{\s}_yK$, $K$ denotes complex conjugation, the term $h_z(\mbk)$ breaks this symmetry.   The lattice has a fourfold symmetry which will be apparent in all quantities to be computed in the following.
\end{widetext}

The noninteracting Chern number of band $\al$ is given by the winding number of $\mbh_\mbk$,\cite{bernevigSCI06,TretiakovPRB07} 
\begin{align}
\nu_\al&={1\over2\pi}\int d^2k {\al\over2}\frac{ \mbf{h}_\mbk\cdot\p_x\mbf{h}_\mbk\times\p_y\mbf{h}_\mbk}{|\mbf{h}_\mbk|^3}\nn
&=-{\al\over2}\int {d^2k\over2\pi}\frac{\p(\cos\theta_\mbk,\varphi_\mbk)}{\p (k_x , k_y)} \nn
&={\al\over2}\sum_i {p(\mbk_i)c(\mbk_i)}\,.
\label{nu}
\end{align}
where $(\theta_\mbk,\varphi_\mbk)$ are the spherical angles of $\mbh_\mbk$,  $c(\mbk_i)$ the chirality, defined by the sense of rotation of $\mbh_\mbk$ as one encircles the vortex center counterclockwise  ($c(\mbk_i)>0$ for counterclockwise), and $p(\mbk_i)= {\rm sgn}\, h_z(\mbk_i)$ is the polarity, which for $B\neq0$ depends on both the sign and magnitude of $\D$.   Since the integrand in Eq.~\eqref{nu} is the Berry curvature of the spinor Bloch states, $\nu_\al$ can be interpreted as the total Berry flux modulo $2\pi$ entering into band $\al$ and is concentrated at $\mbk_i$ with $\sim\pi$ fluxes.  Thus, the Chern number is determined by the sign of the Berry curvature at these points.

The Bloch states at $\mathbf{k}_i$ are eigenstates of $\sigma_z$ with eigenvalues $\alpha p(\mathbf{k}_i)$, a consequence of $\hath(\mathbf{k}_i)$ having 2D spin rotational invariance about the $z$ axis.  These eigenvalues determine the sign of the topological charge $p(\mathbf{k}_i)c(\mathbf{k}_i)/2$ of the quasimomentum space vortices, which  occur in pairs of opposite chirality.\footnote{This is a general property of 2D band structure, according to the fermion doubling theorem.\cite{nielsenNPB81b}}  Thus, when $h_z(\mathbf{k})$ has uniform sign, which occurs for $|\Delta|>8B$,  the Chern numbers vanish due to cancellations among topological charges.  When $|\D|<8B$, the Chern numbers are $\pm1$ and is determined by the $\{(0,\pi),(\pi,0)\}$ topological charges.  The sum of topological charges as a function of $(B,\Delta)$ is shown in Table \ref{topcharge}(right).

In the Mott insulator, the SO texture and corresponding  Chern numbers are determined from the single particle Green function, $\hatG(\w,\mbk)$, which, in the strong coupling perturbation theory to be presented in the following,  will inherit the $\mbk$ space symmetries of $\hath_\mbk$.  For the homogeneous ferromagnetism considered here, the in--plane textures will remain unchanged.  Therefore, the interacting Chern numbers are again determined by the polarities, given by sgn $[{\rm tr}[\hat{\s}_z\hatG(\w,\mbk_i)]]$.    Furthermore, it is clear the leading order term in perturbation theory suffices for the purpose of determining the sign, and hence the Chern numbers.  

It is clear from the discussion above that the topological state defined by a nonzero Chern numbers requires a $\mbk$ dependent SO gap,  provided in the noninteracting case by the $\mcal{T}$ symmetry breaking coupling $B$.    In the bosonic Mott insulator, such a gap can be generated by ferromagnetism which spontaneously breaks $\mcal{T}$ symmetry, thus generating a Chern number which is generally different than the noninteracting one.  To illustrate this effect, we will show in the following that interactions in Mott insulating phase can generate the requisite gap even when $B=0$ but $D\neq0$, even with \emph{homogeneous} ferromagnetism, an effect which cannot be captured in a mean field decoupling of the interaction, as for example, in Ref.~[\onlinecite{raghuPRL08}].  

\begin{figure}[t]
\begin{center}
\begin{tabular}{cc}
\includegraphics[width=.5\linewidth]{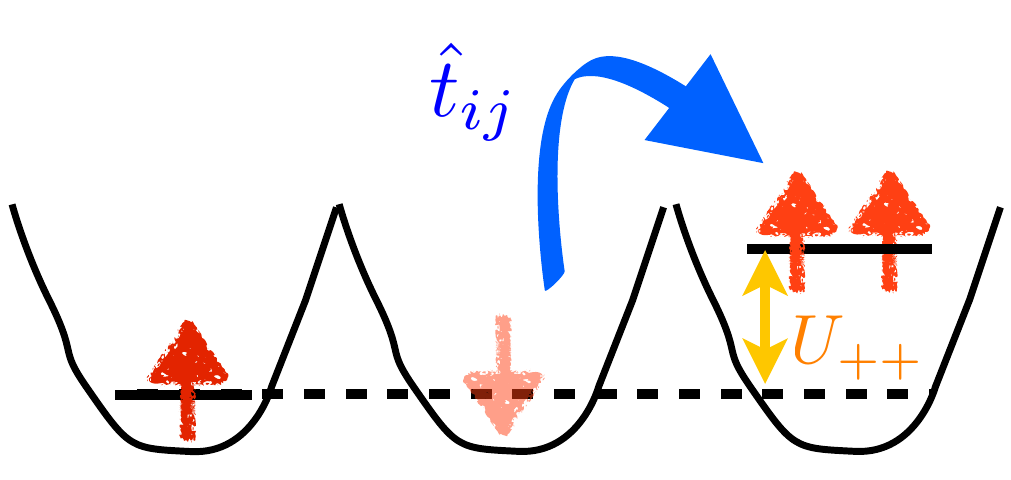}&
\includegraphics[width=.5\linewidth]{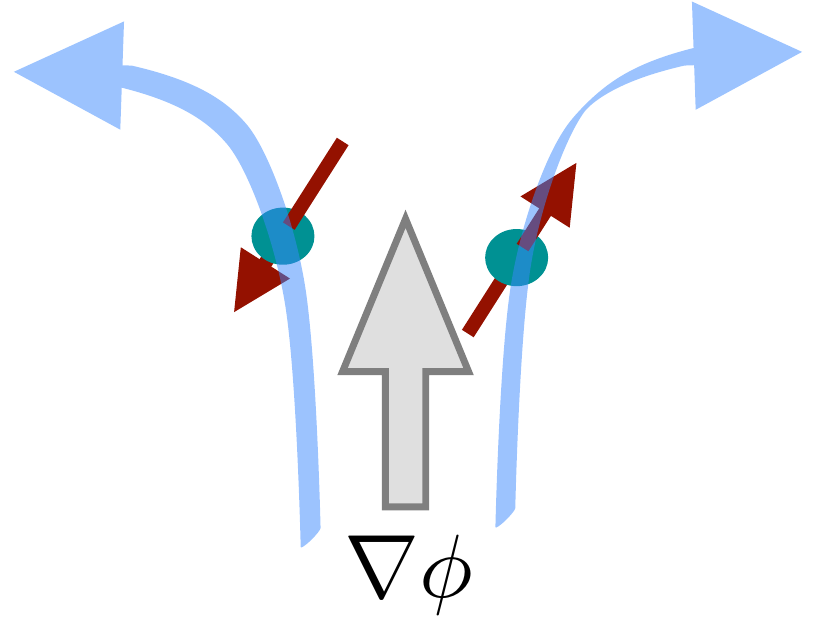}\\
(a)&(b)\\
\end{tabular}
\caption{(Color Online)
Illustration of: (a) Particle/hole excitations in the Bose Hubbard model. (b){ Spin--dependent anomalous velocities. The opposite orientation of the spins indicate the two spin--orbit bands.}}
\label{texture}
\end{center}
\end{figure}
\section{general properties of the Bulk quasiparticle propagator}
\label{Gfcn}
In this section, we summarize some general properties of the quasiparticle propagator in the Mott insulator, express the particle and spin density distributions in terms of the propagator, and establish the notation we use in the following sections.  The results obtained in this section will be used to derive an expression for the interacting Chern numbers in terms of the quasihole Berry curvatures in section \ref{berrychern}. 
\subsection{Spectral representation}
%\footnote{Since $a_{i\al}$ is in the Wannier basis, the real space propagator is given by
%\[
%G_{\al\be}(\mbr-\mbr')
%\<\phi_\al(\mbr)\phi^\dag_\be(\mbr')\>
%=\sum_{\mbk}e^{i\mbk\cdot(\mbr-\mbr')}u_{\mbk\al}(\mbr)u^\dag_{\mbk\be}(\mbr')G_{\al\be}(\w,\mbk)
%\]}
The quasimomentum space, zero temperature, time-ordered propagator is defined by
\ben
G_{\al\beta}(\w,\mbk)=-i\int dt\,e^{i\w t}\<{T} a_{\mbk\al}(t)a_{\mbk\beta}^\dag(0)\>\,,
\label{G}
\een
where $\<{T}\ldots\>$ denotes time-ordered expectation values in the ground state and $\al,\beta$ denote spin indices in the ``lab" frame.  Considering in this section only bulk single particle excitations, which are particle/hole excitations [illustrated in \fig{texture}(a)] the spectral representation of the propagator,  the quasiparticle poles read\footnote{It may be instructive to compare to the Fermi liquid, where the the particle and hole terms can be combined because there is no minus sign between them [See for example, Ref.~[\onlinecite{landauSP2}]
\[
\hat{G}(\w,\mbk)=\sum_{s}\frac{\vec{\chi}_{\mbk s}\vec{\chi}_{\mbk s}^{\dag}}{\omega-\e_{\mbk s}+i\, {\rm sgn} (\w)0^+}\,.
\]
}
\ben
\hat{G}(\w,\mbk)=\sum_{s,n}\frac{ Z_{\ksn}\vec{\chi}_\ksn\vec{\chi}_\ksn^{\dag}}{\omega-\e_\ksn+in0^+}\,,
\label{qpprop}
\een
where $n=\pm$ labels quasiparticle $(+)$ and quasi-hole $(-)$ bands, $s=\pm$ labels the spin-orbit bands, and $\e_\ksn$ are excitation energies relative to the chemical potential which satisfy $\e_{\mbk s+}>0$ and $\e_{\mbk s-}<0$, provided we are in the Mott insulating phase. The numerator of Eq.~\eqref{qpprop} defines normalized spinor wave functions
\[
\sqrt{ Z_{\mbk s+}} \vec{\chi}_{\mbk s+}=\<\mbk s +|\vec{a}^\dag_{\mbk}|\W\>\,,
\sqrt{|Z_{\mbk s -}|}\vec{\chi}_{\mbk s-}=\<{\mbk s-}|\vec{a}_{\mbk}|\W\>\,,
\]
where $|\W\>$ denotes the ground state and the spectral weights are given by
\[
Z_{\mbk s\pm}=\pm\sum_\al|\<{\mbk s\pm}|{a}_{\mbk\al}|\W\>|^2\,,
\]
which measures the probability of creating quasiparticles or holes by adding or removing an atom.  
We can define the spinors above  by ($\mbk$ space) unitary transformations of reference spinors  $\vec{\eta}_s$ with spin quantization axis along the $z$ direction
\[
\vec{\chi}_\ksn=\hatU_{\mbk s n}\vec{\eta}_s\,,\quad \vec{\eta}_\pm=\{(1,0),(0,1)\}\,.
\]
that satisfy
 \[\hatU_\ksn^\dag(\bfhat{f}_\ksn\cdot\hat{\bm{\s}})\hatU_\ksn=\s_z\,,
 \]
where $\bfhat{f}_\ksn$ is a unit $\mbk$ dependent vector in direction of the spinors' quantization axis and which defines the quasiparticle SO field.  The numerators in Eq.~\eqref{qpprop} are projection operators along the SO field, so that the propagator near its poles can be written as
\ben
\lim_{\w\to\e_\ksn}\hat{G}(\w,\mbk)=Z_\ksn\frac{(1+s\bfhat{f}_\ksn\cdot\hat{\bm{\s}})/2}{\omega-\e_\ksn+in0^+}
\label{G2}
\een
 %For each $n=\pm$ these spinors are orthonormal and complete,
% One can define projection operators by 
%\[\vec{\chi}_\ksn\vec{\chi}_\ksn^{\dag}=\frac{1\pm\bfhat{f}_\ksn\cdot\hat{\bm{\s}}}{2}\,,\]
The unitary transformations define above separately diagonalizes the particle and hole parts of the propagator Eq.~\eqref{qpprop}, and they are the on-shell versions of the ones which diagonalizes $\hatG(\w,\mbk)$: $\hatU_\ksn=\hatU(\w\to\e_\ksn,\mbk)$.
%  hey  can be expressed in the Euler-angle  parametrization,
% \ben
%\hat{U}(\theta,\varphi,\gamma)=e^{-i\varphi\hat{\s}_z}e^{-i \theta \hat{\s}_y}e^{-i \gamma\hat{\s}_z}= \left( \begin{array}{cc} 
%  e^ {-i\frac{\varphi+\gamma}{2}} \cos {\theta\over2} &    -e^ {-i\frac{\varphi-\gamma}{2}} \sin {\theta\over2}  \nn
%  e^ {i\frac{\varphi-\gamma}{2}} \sin {\theta\over2}  &     e^ {i\frac{\varphi+\gamma}{2}} \cos {\theta\over2}
%  \end{array} \right )
%  \label{U}
%  \een
The particle/hole parts of the diagonalized propagator,  $G_{ss'}(\w,\mbk)\equiv[\hat{U}^\dag(\w,\mbk)\hat{G}(\w,\mbk)\hat{U}(\w,\mbk)]_{ss'}$ are given by
\ben
\lim_{\w\to\e_\ksn} G_{ss'}(\w,\mbk)\sim\de_{ss'}{Z_\ksn\over\w-\e_\ksn+in 0^+}\,.
 \label{g}
\een
The same information is contained in the spectral function, which for bosons is defined as
\[
A_{\al\be}(\w,\mbk)=\int dt\,d\mbr\,e^{-i\mbk\cdot\mbr+i\w t}\<[a_\al(\mbr,t), a_\be^\dag(0,0)]\>\,.
\]
Integration over frequency gives the equal time commutator which leads to the sum rule (for each $\mbk$)
\ben
\int {d\w\over2\pi} A_{\al\beta}(\w,\mbk)=\<[a_{\mbk\al}, a_{\mbk\be}^\dag]\>=\de_{\al\be}\,.
\label{sumrule}
\een
It is related to the imaginary time propagator by
\ben
\hat{G}(
i\w_n)=\int {d\w'\over2\pi}\, \frac{\hat{A}(\omega',\mbk)}{i\w_n-\w'}\,,
\een
or equivalently, 
\ben
 {\hat{A}(\omega,\mbk)}=i[\hatG(\w+i0^+,\mbk)-\hatG(\w-i0^+,\mbk)]\,.
 \label{spectral}
\een
Similarly, its eigenvalues are given by $A_{ss'}(\w,\mbk)\equiv[\hat{U}^\dag(\w,\mbk)\hatA(\w,\mbk)\hat{U}(\w,\mbk)]_{ss'}$

So far, we have made general, formal deductions based the spectral representation.  In our perturbation theory, we will first calculate the quasimomentum space inverse propagator, which can generally be written as\footnote{Note that in our perturbation theory [cf.~section \ref{pert}],  the self energy as defined here has a zeroth order contribution from the unperturbed onsite propagator, in contrast to the weak coupling perturbation theory, where the self energy contains only perturbative terms.}
\ben
\hatG^{-1}(\w,\mbk)=\w+\mu-\hat{h}(\mbk)-\hat{\Sigma}(\w,\mbk)\,.
\label{invG}
\een
When the self energy is Hermitian, which is the case that we will encounter,\footnote{We find that the imaginary parts are negligibly small.} the eigenvalues and eigenspinors $\vec{\chi}_s(\w,\mbk)$ are given by
\[
\hatG^{-1}(\w,\mbk)\vec{\chi}_s(\w,\mbk)=G^{-1}_{s}(\w,\mbk)\vec{\chi}_s(\w,\mbk)\,,
\]
where $G^{-1}_{s}(\w,\mbk)=\w-H_s(\w,\mbk)\,,$ $H_s$ being eigenvalues of $\hat{H}(\w,\mbk)=\hat{h}(\mbk)+\hat{\Sigma}(\w,\mbk)\,.$

Since quasiparticle energies follows from solutions of the equation $G^{-1}_{s}(\w=\e_\ksn,\mbk)=0$,  the quasiparticle spinor wavefunctions,  $\vec{\chi}_\ksn=\vec{\chi}_s(\w=\e_\ksn,\mbk)$, are the on-shell versions of the $(\w,\mbk)$ eigenspinors with zero eigenvalue.   Near the zeros, {which in the Mott insulator has two roots corresponding to particles and holes,} we have
\begin{align*}
\lim_{\w\to\e_\ksn}G^{-1}_{s}(\w,\mbk)&=Z^{-1}_\ksn(\w-\e_\ksn)\,,\nn
Z^{-1}_\ksn&={\p G^{-1}_s\over\p\w}\Big|_{\w=\e_\ksn}=1-{\p \Sigma_s \over\p\w}\Big|_{\w=\e_\ksn}\,,
\end{align*}
where $\Sigma_s$ denotes eigenvalues of $\hat{\Sigma}(\w,\mbk)$.

It will furthermore be convenient to introduce the four vector parametrization
\begin{align}
-\hat{G}^{-1}(\w,\mbk)
&=d(\w,\mbk)+\mbf{d}(\w,\mbk)\cdot\hat{\bm{\s}}\nn
&=-\sum_sG^{-1}_s(\w,\mbk)P_s(\w,\mbk)\,,
\label{d}
\end{align}
where the eigenvalues and projection operators are given by
\ben
-G^{-1}_s(\w,\mbk)=d+s|\mbd|\,,\,P_s={1+s\hat{\mbd}\cdot\hat{\bm{\s}}\over2}\,,
\label{Gs}
\een
where $\hat{\mbd}=\mbd/|\mbd|$.  Then we have,
\ben
Z_{\ksn}^{-1}=-{\p(d+s|\mbf{d}|)\over\p\w}{\Big\lvert}_{\omega=\e_{\ksn}}\,,
\label{Z}
\een
and the quasiparticle SO fields are $\mbf{f}_\ksn=\mbf{d}(\w=\e_\ksn,\mbk)$.

\subsection{Particle and spin density}
The ground state momentum space distributions of atomic particle and spin density are defined by
\begin{align*}
\left(\begin{array}{c}n_\mbk\\\mbf{s}_\mbk\end{array}\right)&=\< \va^\dag_\mbk\left(\begin{array}{c}1\\\hat{\bm{\s}}\end{array}\right)
\va_\mbk\>\nn
&=i\oint_{{\rm Im}\w>0} d\w\, e^{i\w0^+}{\rm tr}\left[
\left(\begin{array}{c}1\\\hat{\bm{\s}}\end{array}\right)
\hat{G}(\w,\mbk)\right]\,,
\end{align*}
where in the first equality, the correlation functions are for equal time, and in the second, they are expressed in terms of the time-ordered propagator by maintaining the correct order of operators with the limiting procedure $t\to-0^+$ in Eq.~\eqref{G}.   Defining the band $s$ density distributions by $n_{\mbk s}=\langle a^\dag_{\mbk s} a_{\mbk s}\rangle$ and using the form of the propagator from  Eq.~\eqref{G2}, we have 
\[
n_{\mbk s}=-{Z_{\mbk s-}}\,,\quad 
n_\mbk=\sum_{s}n_{\mbk s}\, \quad
\mbf{s}_\mbk=\sum_s sn_{\mbk s}\bfhat{f}_{\mbk s-}
\]
As expected, the spin density is simply the difference of the spin $s$ particle density along the quasiparticle spin quantization axis. 

One can measure these distributions using phase--contrast imaging following a time of flight expansion.\cite{sauPRB11}   But in such an experiment, one measures the real (not quasi) momentum distribution.  Recalling that the Bloch functions, which will be the basis function in our perturbation theory presented in section \ref{pert}, have the momentum space expansion, 
\[ 
\psi_{\mbk\al}(\mbr)=\sum_{\mbK\in(2\pi n/a)\{\bfhat{x},\bfhat{y}\}}c_{\mbk-\mbK,\al}e^{i(\mbk-\mbK)\cdot\mbr}\,,\quad n \in \mathbb{Z}\,,
\]
where $\mbK$ are the reciprocal lattice vectors for a square lattice, $c_{\mbk-\mbK,\al}$ are expansion coefficients, and denoting the second quantized (real) momentum-space operators by $c_{\mbk\al}$, it follows that
\[
\left(\begin{array}{c}n_\mbk\\\mbf{s}_\mbk\end{array}\right)=\sum_{\mbK ,s}\langle \vec{c}^\dag_{\mbk+\mbK, s} \left(\begin{array}{c}1\\\pauli\end{array}\right)
\vec{c}_{\mbk+\mbK, s}\rangle\,,
\]
where the right hand side are the distributions measured in experiments.

\section{Berry curvature and chern numbers \label{berrychern}
}
The topology of the ground state wave function can be characterized by the Chern and spin Chern number, denoted by  $C_0,C_z$, which are topological invariants that can be expressed in terms of the single particle propagator, given in two spatial dimensions by 
\begin{align}
C_a=-{\e^{ij}\over8\pi^2}\int_{{\rm Im}\w>0}&e^{i\w0^+} d\w\int d\mbk \,\nn
& {\rm tr}[\hsig_a\p_\w \hat{G}\p_i \hat{G}^{-1}\hatG\p_j \hatG^{-1}]\,,
\label{Ca}
\end{align}
where here and below $i=(k_x,k_y)$ and $0^+$ denotes a positive infinitesimal, $a=0$ or $z$ with $\hat{\s}_0=\hat{1}$, and $\hatG(\w,\mbk)$ is defined in Eq.~\eqref{G}.  The Hall conductivity is given by $\s_H=C_0/h$,\cite{ishikawaNPB87} and is equivalent to the Chern number defined by the many body wave function.\cite{wangPRL10to}  If spin is conserved,    $C_z$ is related to the spin Hall conductivity.\cite{yoshidaPRB12} 

%This expression is a topological invariant in frequency and momentum space\cite{volovikBook03} called the Chern number for an interacting system and is related to the number of edge states, and a nonzero integer value defines QAH phase.\cite{wangPRL10to,qiPRB06} 

 The frequency integration in Eq.~\eqref{Ca} picks up many body excitations corresponding to  singularities in $\w$ in the upper complex plane.  Although it is possible to evaluate this expression directly in terms of the Green functions,  when there are only quasiparticle poles (as in the case for the approximation taken in this paper) it is useful to express $C_a$ in terms of the quasiparticle Berry curvature.\cite{shindouPRL06}  To this end,  we define the $(\w,\mbk)$ space matrix gauge field $\hatA_\mu\equiv i\hat{U}^\dag\p_\mu\hat{U}$, which arises in the integrand of Eq.~\eqref{Ca} when the Green function is rotated in spin space to the SO basis.\footnote{They are well defined as long as the SO fields are nonvanishing, tr$[\hsig_i\hatG(\w,\mbk)]\neq0$ (which is always true with a finite $\D$), so that $\hat U(\w,\mbk)$ and hence $\hatA_\mu(\w,\mbk)$ is nonsingular. 
% {They singular $(\w,\mbk)$ points where the bands cross, where $\hat{G}\propto\hat{1}$.We will not encounter these points because we always consider a finite $\D$. 
 }   Denoting the diagonalized Green function by $\hatG_d$, we find  
\begin{align}
C_a&={1\over4\pi}\int_{\rm BZ}{d\mbk} \oint {d\w\over2\pi i} e^{i\w0^+}\nn
&\,2\e^{ij}{\rm tr}[\hsig_a(\hatG_d^{-1}\p_\w \hatG_d\hatA_i\hatA_j+\hatG_d^{-1}\p_i \hatG_d\hatA_j\hatA_\w)]\,.
\label{Ca1}
\end{align}
Then, the only $\w$ singularities come from $\hatG_d$,
and their contributions read [cf.~Eq.~\eqref{g}]
\[\lim_{\w\to\e_\ksn}
\left(\begin{array}{c}
\hatG_d^{-1} \p_\w \hatG_d\\
\hatG_d^{-1} \p_\mbk \hatG_d\end{array}\right)
\sim-{\hat{1} \over\w-\e_\ksn+in0^+}\, 
\left(\begin{array}{c} 1\\\mbv_{\ksn}\end{array}\right)\,,
\]
where $\hat{1}$ is the $2\times2$ unit matrix, $n=\pm$ for particles/holes and $\mbv_{\ksn}=\p_\mbk\e_{\ksn}$. We thus pick up only hole contributions with poles at $\w=\e_{\mbk s-}+i0^+$.  Note that in the terms above we have omitted the terms containing it $\sim\p _\mbk Z_\ksn/Z_\ksn$ which is finite and vanishes in the $\w$ integration in Eq.~\eqref{Ca1}.

Next, we define  Berry gauge field by the diagonal part $\mcal{A}^s_{\mu}\equiv[\hatA_\mu]_{ss}$ and the Berry field strength tensor by the diagonal part of the commutator that  appears in the integrand of Eq.~\eqref{Ca1} 
\[
\mcal{F}^s_{\mu\nu}\equiv i([\hatA_\mu,\hatA_\nu])_{ss}=\p_\mu\mcal{A}^s_{\nu}-\p_\nu\mcal{A}^s_{\mu}\,,
\] 
with $\mu,\nu=(\w,k_x,k_y)$.  The second equality follows from the fact that $\hatA_\mu$ is ``pure gauge,'' so that its non-abelian field strength is always zero.  The associated Berry electric and magnetic fields $\bm{\mcal{E}}_s(\w,\mbk)$, $\mcal{B}_s^z(\w,\mbk)$, respectively, are defined given by $(\mcal{E}^i_{s},\mcal{B}^z_{s})=(\mcal{F}^s_{\w i},\mcal{F}^s_{xy})$. 

Picking up the contributions from the poles, $C_a$ can be expressed in terms of band $s$ hole Chern numbers $C_s$
\begin{align}
\left(\begin{array}{c}C_0\\C_z\end{array}\right)=
\left(\begin{array}{c}C_++C_-\\C_+-C_-\end{array}\right)\,,\,\,\,
 C_s=-\int_{\rm BZ}{d\mbk\over2\pi}\,\mcal{C}_{\mbk s-}\,,
\label{Cs}
\end{align}
where $\mcal{C}_\ksn$ is the $\mbk$ space Berry curvature, which can be expressed in terms of the on-shell Berry electromagnetic fields, 
\begin{align}
{\mcal{C}}_\ksn&=\mcal{B}^z_\ksn+\mbf{v}_\ksn\times\bm{\mcal{E}}_\ksn\,,\\
(\bm{\mcal{E}}_\ksn,\mcal{B}^z_\ksn)&=
\left(\bm{\mcal{E}}_s(\w,\mbk),\mcal{B}_s^z(\w,\mbk)\right)\Big|_{\w=\e_\ksn}\,. \nonumber
\label{C}
\end{align}
The electric field contribution $\bm{	\mcal{E}}_\ksn$ to the total Berry curvature is strictly an interaction effect which requires nontrivial frequency dependence in the matrix structure of $\hatG(\w,\mbk)$ and thus $\hatU(\w,\mbk)$.  In the absence of interactions, $\bm{\mcal{E}}_\ksn=0$ and $\mcal{B}^z_{\mbk s}=\mcal{B}^z_s(\w,\mbk)$ is independent of $\w$, and $C_s$ reduces to the noninteracting  Chern number, which is a well known integer topological invariant.  In the presence of interactions, nontrivial $\w$ dependence modifies both electric and magnetic contributions, but in such a way that the integral of Berry curvature in Eq.~\eqref{Cs} remains an integer.
This fact can be made manifest by expressing $\mcal{C}_\ksn$ as the curl of the on--shell gauge fields
\[
\mcal{C}_\ksn=\p_x{\mcal{A}}^{y}_\ksn-\p_y{\mcal{A}}^{x}_\ksn\,,\]
which are defined in terms of on--shell matrix rotations [cf. paragraph below Eq.~\eqref{G2}]  
\[
{\bm{\mcal{A}}}_{\ksn}\equiv i[\hat{U}_\ksn^\dag\p_\mbk\hat{U}_\ksn]_{ss}=[\mcal{A}^s_{\w}\mbv_\ksn +{\mcal{A}}^s_{\mbk}]|_{\w=\e_\ksn}\,.
\] 

In terms of the $\mbd$ vector parametrization in Eq.~\eqref{d}, the on--shell Berry field strength and its associate electromagnetic fields are given by
\[
\mcal{F}^s_{\mu\nu}({\w=\e_\ksn},\mbk)={s\over2}\mbf{\hat{d}}\cdot\p_\mu\mbf{\hat{d}}\times\p_\nu\mbf{\hat{d}} \Big|_{\w=\e_\ksn}\,.
\]
In terms of the quasiparticle spin-orbit field in $\mbk$ space,  it is readily verified that  
\ben
\mcal{C}_\ksn={s\over2}{\bfhat{f}_\ksn\cdot\p_x\bfhat{f}_\ksn\times\p_y\bfhat{f}}_\ksn\,,
\label{C}
\een
and thus the hole band Chern numbers $C_s$ are the integer winding number of the hole band SO fields $\mbf{f}_{\mbk sn}$. 

While the Chern numbers describe the global topology of the quasiparticle Hamiltonians, the Berry curvature describes its local geometry, and furthermore, causes in the quasiparticle wave packet motion an anomalous velocity transverse to external forces. The semiclassical equations of motion are given by
\begin{align}
\dot{\mbr}_{sn}&=\mbv_\ksn+\mcal{C}_\ksn(-\del\phi\times\mbf{\hat{z}})\nn
\dot{\mbk}_{sn}&=-\del\phi
\label{sEOM}
\end{align}
where $\phi$ is an external potential.  Thus, the Berry curvature can be measured experimentally by detecting single particle wave packet dynamics under external forces, which can be applied, for example, by ramping the optical lattice potential,\cite{pricePRA12,liuCM13} or simply be due to the force of gravity.\cite{zhuPRL06}  In addition, $\phi$ always includes a trapping potential, and the corresponding anomalous velocity produces an edge current.  \cite{matsumotoPRL11} This band dependent velocity is illustrated in Fig.~\ref{texture}(b).

In practice, a typical cold atoms experiments would detect not a single wave packet but the phase space distribution function, which satisfies a semiclassical Boltzmann equation with advective terms 
that read 
\[
(\p_t+\dot{\mbf{r}}_{sn}\cdot\p_\mbr +\dot{\mbf{k}}_{sn}\cdot\p_\mbk) f_\ksn(\mathbf{r},t)=\ldots
\]
where $\dot{\mbr}_{sn},\dot{\mbk}_{sn}$ are given in Eq~\eqref{sEOM},  $f_\ksn$ is the semiclassical distribution function, and $\ldots$ denotes collisional and other possible terms, but the rigorous derivation of the Boltzmann equation is beyond the scope of this paper. However, because we are considering an interacting system, the semiclassical Boltzmann equation should be properly derived from a quantum kinetic equation, which can be based on a density matrix or nonequilibrium Green function formalism.\cite{wongPRB11,shindouPRL06}   The single particle equations of motion,  $\dot{\mbr}_{sn},\dot{\mbk}_{sn}$, can then be inferred  from the advective terms.

We  note here that the Berry curvature and Chern numbers can change from zero in the noninteracting limit to nonzero in the strongly interacting limit studied in this paper.  An example of the purely interaction generated Berry curvature (coming from the Berry electric field) is given  by the experimentally realized  SO coupling  studied in Ref.~[\onlinecite{wongPRL13}]. In this paper, we will consider the case  the noninteracting Chern numbers are zero when $B=0$, while the interacting Chern numbers are nonzero for a range of values of the hopping parameters $D,\D$, as shown in the phase diagram in  \fig{chern}(d).

When the Chern numbers are nonzero, we expect gapless edges states.  However, in an experiment, the confining potential effectively raises the chemical potential of at the edge of the system, resulting a superfluid phase at the edge of the Mott insulator, while our formalism is only valid in the Mott insulating phase [cf. appendix \ref{edge}].

\begin{widetext}
\section{Strong coupling perturbation theory}
\label{pert}
In this section, we present the computation of $\hat{G}^{-1}(\w,\mbk)$ and related quantities in the Mott insulator.   We use a perturbation theory in which the hopping Hamiltonian $V$ is the perturbation to the interaction Hamiltonian $H_0$. \cite{fisherPRB89,oostenPRA01}  We will compute the imaginary time ($\tau=it$) propagator 
 \[
  G_{i\al,j\be}(\tau-\tau')=-\<Ta_{i\al}(\tau)a^\dag_{j\be}(\tau)\>\,,
 \]
 which, to zeroth order, is given by onsite propagator, which has the path integral representation (here and below we set $\hbar=1$)
\ben
-\<{T}a_{i\al}(\tau')a^\dag_{j\beta}(\tau)\>_0=-\int\mcal{D}\va\mcal{D}\va^*\, a_{i\al}(\tau')a^*_{i\beta}(\tau)e^{-S_0[\va]}\equiv\de_{ij}\hat{g}_i(\tau,\tau')\,,
\een 
which is local because $H_0$ contains only onsite interactions.  Next, consider the grand canonical partition function with external sources $\vJ$, 
\begin{align}
Z[\vJ,\vJ^*]&=\int \mcal{D}\va \mcal{D}\va^*\,  \exp\left[-S_0[\va]+\int d\tau\left(\sum_{i,j}-\va_i^*\cdot\hat{t}\cdot\va_j-\sum_i\vJ_i^*\cdot\va_i+\va_i^*\cdot\vJ_i\right)\right]\,;\nn
S_0&=\int d\tau\,\sum_{i} [\va_{i}^*\p_\tau \va_{i}+H_0(\va_i,\va^*_i)]\,.
\end{align}
Using the Hubbard Stratonovich transformation, we decouple the term quadratic in $\va$ by introducing an integration over an auxillary field $\int \mcal{D}\vpsi \mcal{D}\vpsi ^*e^{-S_{\rm hb}}$, thus adding to the action\cite{fisherPRB89}
\[
S_{\rm hb}=-\int d\tau\sum_{i,j}(\va_i^*-\vpsi_i^*-\vJ_{i'}^*\cdot \hat{t}^{-1}_{i'i})\hat{t}_{ij}(\va_j-\vpsi_j-\hat{t}_{jj'}^{-1}\cdot\vJ_{j'})\,,
\]
so that the partition function is proportional to
\begin{align}
Z[\vJ,\vJ^*]
&\propto \int\mcal{D}\vpsi \mcal{D}\vpsi ^* \exp\left[\int d\tau\left(\sum_{i,j}\vpsi_i^*\hat{t}_{ij}\vpsi_j+\vJ_i^*\cdot\hat{t}_{ij}^{-1}\vJ_j
+(\sum_i\vJ_i^*\vpsi_i+\vpsi_i^*\vJ_i)\right)-S_1[\vpsi]\right];\nn
S_1[\vpsi]&=-\ln\int\mcal{D}\va\,\exp\left[-S_0[\va]-\int d\tau \sum_{i,j}\left(\va_i^*\hat{t}_{ij}\vpsi_j+ \vpsi_i^*\hat{t}_{ij}\va_j \right)\right]\,.
\label{S1eq}
\end{align}
\end{widetext}
Next, we define the effective action (landau free energy) for $\vpsi$,
\begin{align}
Z[\vJ,\vJ^*]&\equiv\int\mcal{D}\vpsi \mcal{D}\vpsi^*\exp({-W[\vpsi,\vJ]})\,,\nn
W[\vpsi;\vJ=0]&=-\int d\tau \sum_{i,j}\vpsi_i^*\hat{t}_{ij}\vpsi_j+S_1[\vpsi]\,.
\label{W}
\end{align}
By differentiating with respect to the sources, it follows readily that 
\begin{align}
\<a_{i\al}\>&={1\over Z}\frac{\de Z}{\de J_{i\al}}|_{J=0}=\<\psi_{i\al}\>\,,\nn
\<Ta_{i\al} a_{j\beta}^\dag\>
&={1\over Z}\frac{\de^2Z}{\de J_{i\al} \de J_{j\beta}^*}|_{J=0}=t^{-1}_{j\beta,i\al}+\<T\psi_{i\al}\psi_{j\beta}^\dag\>\,,
\end{align}
or in Fourier modes 
%\[
%-G_{\al\beta}^{(a)}(\mbk)=\<a_{\mbk\al} a_{\mbk\beta}^\dag\>=\sum_{n,ij}e^{-i\mbk\cdot(\mbr_i-\mbr_j)-i\w_n(\tau-\tau')}\<a_{i\al}(\tau) a_{j\beta}^\dag(\tau')\>\,,
%\]
\ben
\hat{G}(i\w,\mbk)=\hat{G}^{(\psi)}(i\w,\mbk)-\hat{h}^{-1}_\mbk\,,
\label{Ga}
\een
where $G,G^{(\psi)}$ denote the atomic and superfluid propagators, respectively.  So far, we have made no approximations, so that Eq.~\eqref{Ga} is a general relation between the atomic and the superfluid propagator.  The superfluid propagator can now be computed perturbatively by making a cumulant expansion of $S_1$ in $\vpsi$.  However, because $H_0$ contains quartic terms, there are no Feynman rules similar to the weak coupling perturbation theory.

We will calculate the propagator in mean field theory by doing a saddle point approximation of the path integral at $\vpsi=0$, since we are in the Mott insulating phase.   This means we keep  $S_1[\psi]$ in Eq.~\eqref{S1eq} up to terms quadratic in $\psi$,\cite{oostenPRA01} 
\begin{align}
W^{(2)}[\vpsi]&=-\int\,d\tau\sum_{i,j}\vpsi_i^\dag\hat{t}_{ij}\vpsi_j+S_1^{(2)}[\vpsi]\nn
&\equiv-\int{d\w d\mbk\over(2\pi)^3}
\vpsi^\dag_{\mbk\w}G^{(\psi)-1}(i\w,\mbk)\vpsi^\dag_{\mbk\w}\,.
\label{W2}
\end{align}
 The details of the computation are given in appendix \ref{S1}.  We find that the superfluid propagator is given by
\ben
G^{(\psi)-1}(i\w,\mbk)=\hat{h}_\mbk-\hat{h}_\mbk\hat{g}(i\w)\hat{h}_\mbk\,,
\label{invGpsi}
\een
whence from the inverse of Eq.~\eqref{Ga} and after performing the wick rotation back to real time, or in frequency space, after taking $i\w\to\w$,\footnote{This is equivalent to computing finite temperature, Matsubara Green function, and then taking the zero temperature limit.} we find
\ben
\hat{G}^{-1}(\omega,\mbk)=\hat{g}^{-1}(\w)-\hat{h}(\mbk)\,.
\label{invG1}
\een
%$\hat{g}^{-1}(\w)$ is the Fourier transform of the onsite propagator, which is local in space and diagonal in spin with components given by 
%\[g_\al(t)=-i\<N_+N_-|\mcal{T}a_{i\al}(t)a^\dag_{i\al}(0)|N_+N_-\>\,,\]
Comparing Eq.~\eqref{invG1} and \eqref{invG}, the self energy is simply given by $\hat{\Sigma}(\w)=\hat{g}^{-1}(\omega)-\w$,  depending only on $\w$.   We also note that although Eq. \eqref{invG1} is first order in hopping, its inverse, the propagator, contains all orders in hopping.  The four vector components, defined in Eq.~\eqref{d}, are given by 
\begin{align}
\mbf{d}(\w,\mbk;N_\al)&=\mbf{h}(\mbk) -(g^{-1})_z(\w;N_\al)\bfhat{z}\,,\nn
d(\w,\mbk;N_\al)&=h(\mbk)-(g^{-1})_0(\w;N_\al)\,, 
\label{da}
\end{align}
 where 
\[
(g^{-1})_{0}={g^{-1}_{+}+g^{-1}_{-}\over2}\,,\quad(g^{-1})_{z}={g^{-1}_{+}-g^{-1}_{-}\over2}\,,
\]
and the on-shell texture is given by 
 \ben
\mbf{f}_\ksn=\mbf{d}(\omega=\e_\ksn,\mbk)=\mbf{h}_\mbk-(g^{-1})_z(\e_\ksn)\bfhat{z}\,.
\label{f}
\een

Since the in-plane texture is unchanged, the in-plane spin density is simplify given by
\ben
s^i_{\mbk}=(n_{\mbk+}-n_{\mbk-}){h^i_\mbk\over|\mbh_\mbk|}\,,\quad i=x,y\,,
\label{spinxy}
\een
which has the same orientation (up to a sign) as the SO texture of the hopping Hamiltonian $\mbk$, but its magnitude is renormalized by the difference in the band $s$ densities.  On the other hand, the $z$ component is shifted by the $\s_z$ component of the inverse propagator. 

The spectral weights are given explicitly by [cf.~Eq.~\eqref{Z}]
 \[
 Z_\ksn^{-1}=\p_\w (g^{-1})_0\Big|_{\w=\e_\ksn}-s\cos\theta_\ksn\p_\w (g^{-1})_z\Big|_{\w=\e_\ksn}\,,\]
 where we define the out--of--plane angle $\cos\theta_\ksn={f^z_\ksn/|\mbf{f}_\ksn|}$.

From Eq.~\eqref{da} the frequency dependent Berry Curvature is given by
\begin{align}
\mcal{B}^z_\ksn&={s\over2}\frac{\mbf{d}\cdot\p_x\mbf{d}\times\p_y\mbf{d}}{|\mbf{d}|^3}\Big|_{\w=\e_\ksn}={s\over2}\frac{\mbf{f}_\ksn\cdot\p_x\mbh_\mbk\times\p_y\mbhk}{|\mbf{f}_\ksn|^3}\,,\nn
\mcal{E}^i_\ksn&={s\over2}\frac{\mbf{d}\cdot\p_\w\mbf{d}\times\p_i\mbf{d}}{|\mbf{d}|^3}\Big|_{\w=\e_\ksn}\nn
&={s\over2}\frac{\p_w(g^{-1})_z}{|\mbf{f}_\ksn|^3}{\mbf{f}_\ksn \cdot\bfhat{z}}\times(\p_x\mbh_\mbk,\p_y\mbh_\mbk)\,.	
\label{EB}
\end{align}

The band $s$ dispersion is computed by setting $G^{-1}_s(\omega=\e_\ksn,\mbk)=0$, which, from Eq.~\eqref{da} Eq.~\eqref{Gs}, can be written as
\ben
(g^{-1})_{0}(\e_\ksn)=h_\mbk+s|\mbf{f}_\ksn|\equiv\til{\e}_\ksn\,.
\label{eqdispe}
\een
By functionally inverting $(g^{-1})_{0}(\e_\ksn)$, it can be expressed implicitly in terms of $\til{\e}_\ksn$, which reduces to the noninteracting hopping energy $\bare_{\mbk}$ if $g^{-1}_z=0$.  To compute the dispersion, on must solve

\begin{align}
h&(\mbk)-(g^{-1})_{0}(\e_{sn},\mu_\pm)\nn
&=-s\sqrt{h_{x}^2(\mbk)+h_{y}^2(\mbk)+[h_z(\mbk)-(g^{-1})_z(\e_{sn},\mu_\pm)]^2}\,,
\label{disp}
\end{align}
for $\e_{sn}$.

\subsection{An approximate expression for $f^z_\ksn$}
\label{fkz}
In the following, we will consider only the hole bands, which are the ones relevant for ground state properties; thus below where equations have two roots, we will consider only the ones corresponding to holes, and for brevity omit the particle/hole index $n$.

The SO gap including interactions is defined by\footnote{In this section, we use the index $\al$ instead of $s$ because we will focus on the high symmetry $\mbk$ points where $\hatG(\w,\mbk)$ is already diagonal, so that the spin states correspond to eigenvalues of $\s_z$.}
\[
-{f}^z_\al(\mbk) =G^{-1}_+(\e_{\mbk \al})-G^{-1}_-(\e_{\mbk \al})=-\al G^{-1}_{-\al}(\e_{\mbk \al})\,,
\]  
which, in our perturbation theory, is given by
\ben
 {f}^z_{\al}(\mbk)=h_z(\mbk)-(g^{-1})_z(\e_{\mbk \al}) \,.
 \label{fz}
 \een
 It determines the Chern numbers in the Mott insulator similar to the noninteracting case discussed in section \ref{hopping}, with the polarity of the vortices now given by
\ben
p_\al(\mbk_i)={\rm sgn} f^z_\al(\mbk_i)\,.
\label{p}
\een
Naively, since $(g^{-1})_z$ is of order $U_{\al\beta}$, one might expect it to be the dominant term.  However, generally in our perturbation theory, the quasiparticle energies are always close to the onsite particle/hole excitation energies, $\xi^{(n)}_\al$, which are the zeros of $({g}^{-1})_z(\w)$, as shown for example in Fig.~\ref{invgfig}(b)(red curve). Therefore, the winding number is determined by the competition between the terms in Eq.~\eqref{fz}, and may \emph{differ from the non-interacting one determined by $h_z(\mbk_i)$}.

An exception occurs for {filling factors $(N,0)$, where one hole band and its corresponding zero in $({g}^{-1})_z(\w)$ is missing, as shown for the $(1,0)$ filling in Fig.~\ref{invgfig}(b)(blue curve).   In this case, $({g}^{-1})_z(\w<0)\sim U$ sets a gap with uniform sign over the whole BZ, thereby making the Chern number zero.} 

\begin{figure}[t]
%\begin{center}
\includegraphics[width=\linewidth]{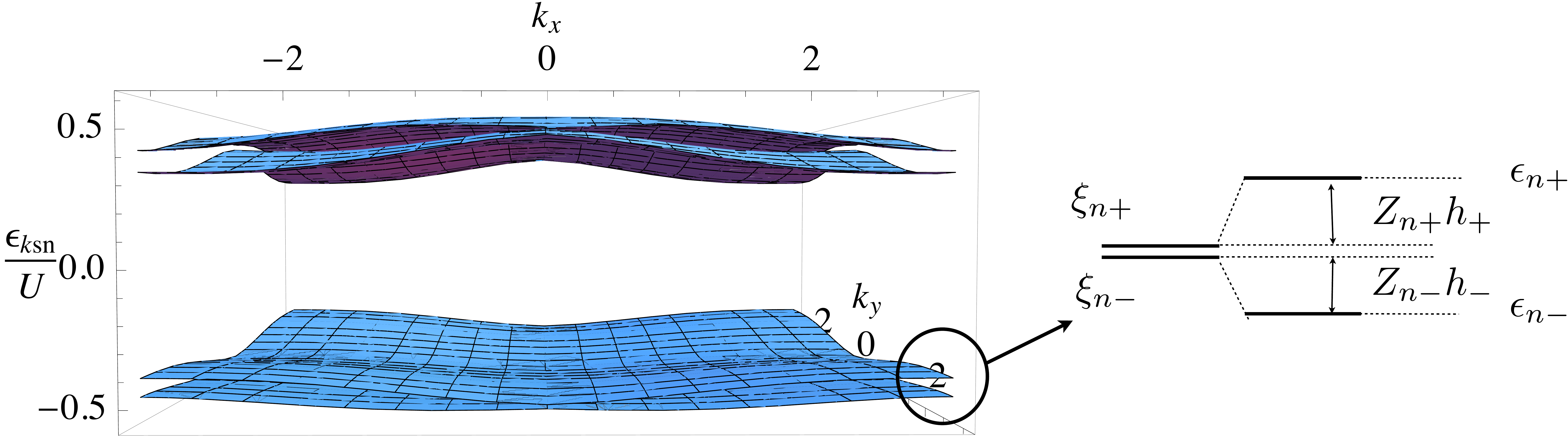}
\caption{(Color Online) (Left) The $(1,2)$ quasiparticle dispersions plotted as functions of wave vectors the Brillouin zone in units of inverse lattice spacing.  (Right) Illustration of the splitting between the spin bands, approximately given by Eq.~\eqref{spingap}.}
\label{SOgap}
%\end{center}
\end{figure}

Next, we illustrate the statements above explicitly by deriving an approximate, analytic expression for SO gaps in Eq.~\eqref{fz} at the vortex cores $\mbk_i$.   At each vortex core, the in-plane texture vanishes, so that Eq.~\eqref{invG1} is diagonal and the energies satisfy 
\ben
{G}^{-1}_{\al}(\e_{\al}(\mbk_i))={g}^{-1}_{\al}(\e_{\al}(\mbk_i))-\e^{(1)}_{\al}(\mbk_i)=0\,,
\label{invGi}
\een
where $\epsilon^{(1)}_{\alpha}(\mathbf{k}_i)$ are the noninteracting energies. The quasiparticle energies are given by the functional inverse of ${g}^{-1}_{\al}$ as a function of $\e^{(1)}_{\al},$\footnote{It is given by an expression similar to Eq.~\eqref{dispN}} which to leading order in $\e^{(1)}_{\al}$ is given
\[
\e_{\al}(\mbk_i)\approx\xi_{\al}+Z_{\al}\e^{(1)}_\al(\mbk_i)\,,
\]
where the spectral weights are simply given by
\[
Z_{\al}^{-1}\equiv\p_\w g_\al^{-1}(\xi_{\al})= -1/N_\al\,.
\]
Thus, the energy splitting between the spin bands is given by
\begin{align}
\de_{\rm SO}(\mbk_i)\equiv\e_{+}(\mbk_i)-\e_{-}(\mbk_i)\approx Z_{+}h_{+}(\mbk_i)-Z_{-}h_{-}(\mbk_i)\,,
\label{spingap}
\end{align}
as illustrated in Fig.~\ref{SOgap}.  Then, linearizing Eq.~\eqref{fz}  with respect to $\de_{\rm SO}$, we find, to leading order
\begin{align}
-f^z_\al(\mbk_i)&\approx-Z_{-\al}^{-1}\de_{\rm SO}(\mbk_i)\nn
&={1\over N_{-\al}}\left[{N_+-N_-\over2}h(\mbk_i)+{N_++N_-\over2}h_z(\mbk_i)\right]\,.\nn
 \label{fz4}
\end{align}
This expression shows that  if $N_+\neq N_-$, even the spin-independent hopping $h(\mbk)$  can lead to a nonzero SO gap, which is not possible in the absence of interactions.  In general, the SO gaps are determined by the competition between $h(\mbk_i)$ and $h_i(\mbk_i)$, thus the Chern numbers can be different than the noninteracting ones.

We close this discussion by noting some simplifications of our results when $h(\mbk_i)=h_z(\mbk_i)=0$,   {In this case, we can generically express $\e_\ksn$ in terms of the noninteracting dispersion, by inverting 
\ben
{\det\hatG^{-1}}=g^{-1}_+(\w)g^{-1}_-(\w)-(\e^{(1)}_{\mbk\al})^2=0\,.
\label{diracdisp} 
\een
So we can find explicit solutions in terms of $|\e^{(1)}_{\mbk\al}|$, which we plot in Fig.~\ref{dispe}(b).   The plot clearly shows that the symmetry of the spectrum between spin $\pm$ bands, which has energies $\pm|\e^{(1)}_{\mbk\al}|$ in the hopping Hamiltonian, is preserved in the Mott insulator.  

This highly symmetric  case occurs at the massless Dirac points of the hopping Hamiltonian, which occurs in our model when $\D=0$ at $(0,\pi)$ and $(\pi,0)$.  Eq.~\eqref{fz4} shows that the SO gap vanishes at these points.   In fact, this is true beyond the approximations taken above because that $\e^{(1)}_{\mbk\al}=0$ at the gap closing points, so that Eq.~\eqref{invGi} reads ${g}^{-1}_\al(\e_{\al}(\mbk_i))=0$, and thus $\e_{\al}(\mbk_i)=\xi_{\al}$ are equal to the onsite hole excitation energies, which are poles of the propagator and thus the zeros of  $g^{-1}_\al(\w)$.  Since we set the chemical potentials so that $\xi_+=\xi_-$,  we also have $G^{-1}_{-\al}(\e_\al(\mbk_i))={g}^{-1}_{-\al}(\e_{\al }(\mbk_i))=0$, from which follows $f^z_\al(\mbk_i)=0$.  Therefore, in our perturbation theory {interactions cannot open a SO gap at a massless Dirac point of the hopping Hamiltonian}.

\section{Onsite propagator\label{onsite}}
The final ingredient needed to obtain results is the onsite propagator.  We will specify the unperturbed eigenstates in the occupation number basis by the number of spin up and spin down particles per site, $(N_+,N_-)$,
then the energies per site are given by 
\ben
E^{(0)}_{N_+ N_-}=\sum_\al\left[{U_{\al}\over 2}N_\al(N_\al-1)-\mu_\al N_\al\right]+U_{+-}N_+N_-\,.
\label{onsiteE}
\een

The unperturbed time ordered correlation function  is local in space and diagonal in spin defined by $g_\al(t)=-i\<N_+N_-|{T}a_{i\al}(t)a^\dag_{i\al}(0)|N_+N_-\>$, and its Fourier transform reads 
 \begin{align}
g_\al(\omega)
&=\frac{1+N_\al}{\omega-\xi_\al^{(+)}+i0^+}-\frac{N_\al}{\omega{-\xi_\al^{(-)}}-i0^+}\,.
\label{onsiteg}
\end{align}
where 
\begin{align*}
\xi_\al^{(+)}&= E_{N_\al+1,N_{-\al}}-E_{N_\al,N_{-\al}}>0\,;  \\
\xi_\al^{(-)}&= E_{N_\al,N_{-\al}}-E_{N_\al-1,N_{-\al}}<0 \,,   
\end{align*}
are the excitation energies, and the inequalities must be satisfied for the system to be in the Mott phase, which restricts the chemical potentials to  
\[\sum_{\be}U_{\al\be}N_\be-1<\mu_\al<\sum_{\be}U_{\al\be}N_\be\,,\]
i.e., they lie in the particle/hole gap.
\begin{figure}[t]
\begin{center}
\begin{tabular}{ccc}
\includegraphics[width=.5\linewidth]{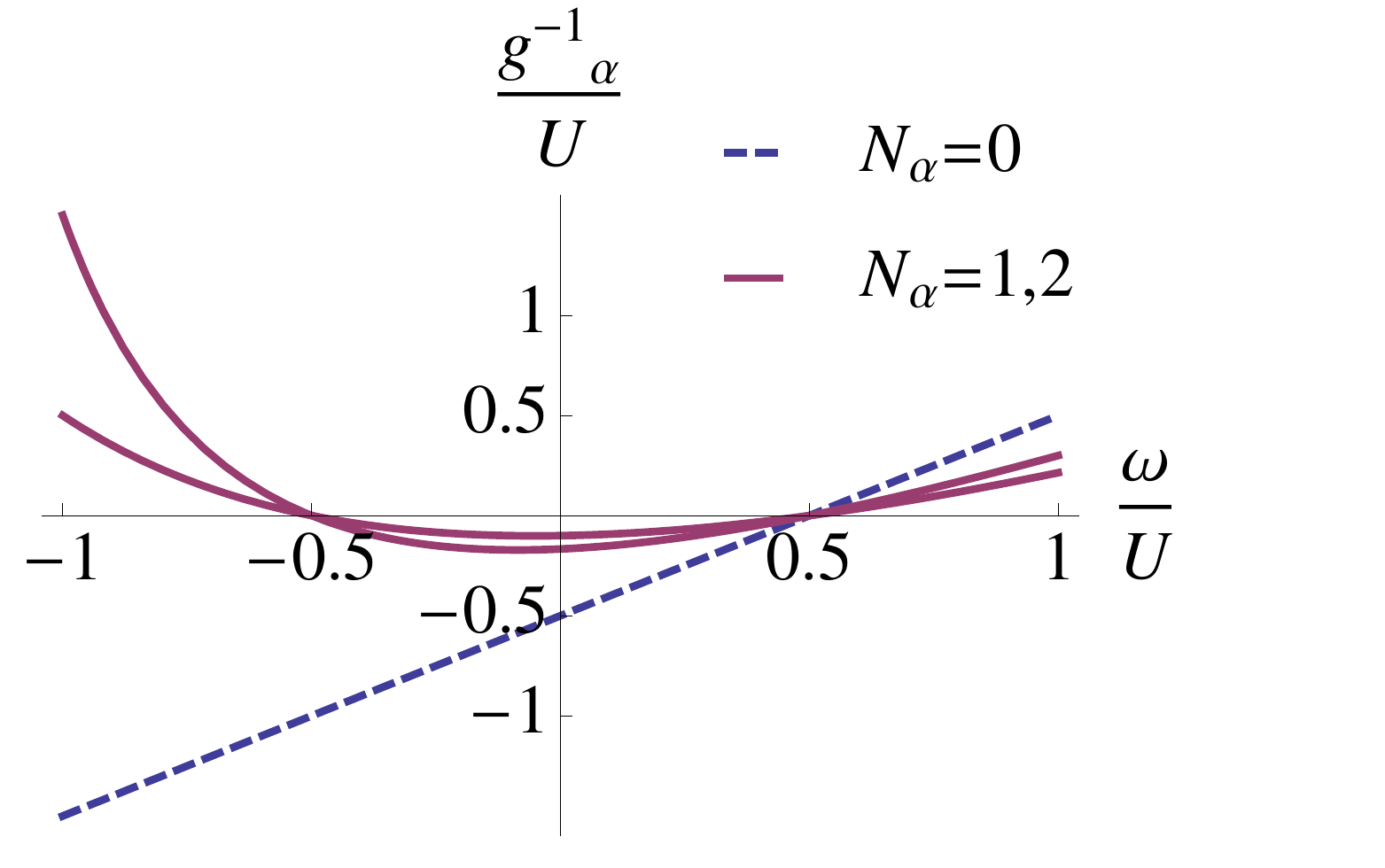}&
\includegraphics[width=.5\linewidth]{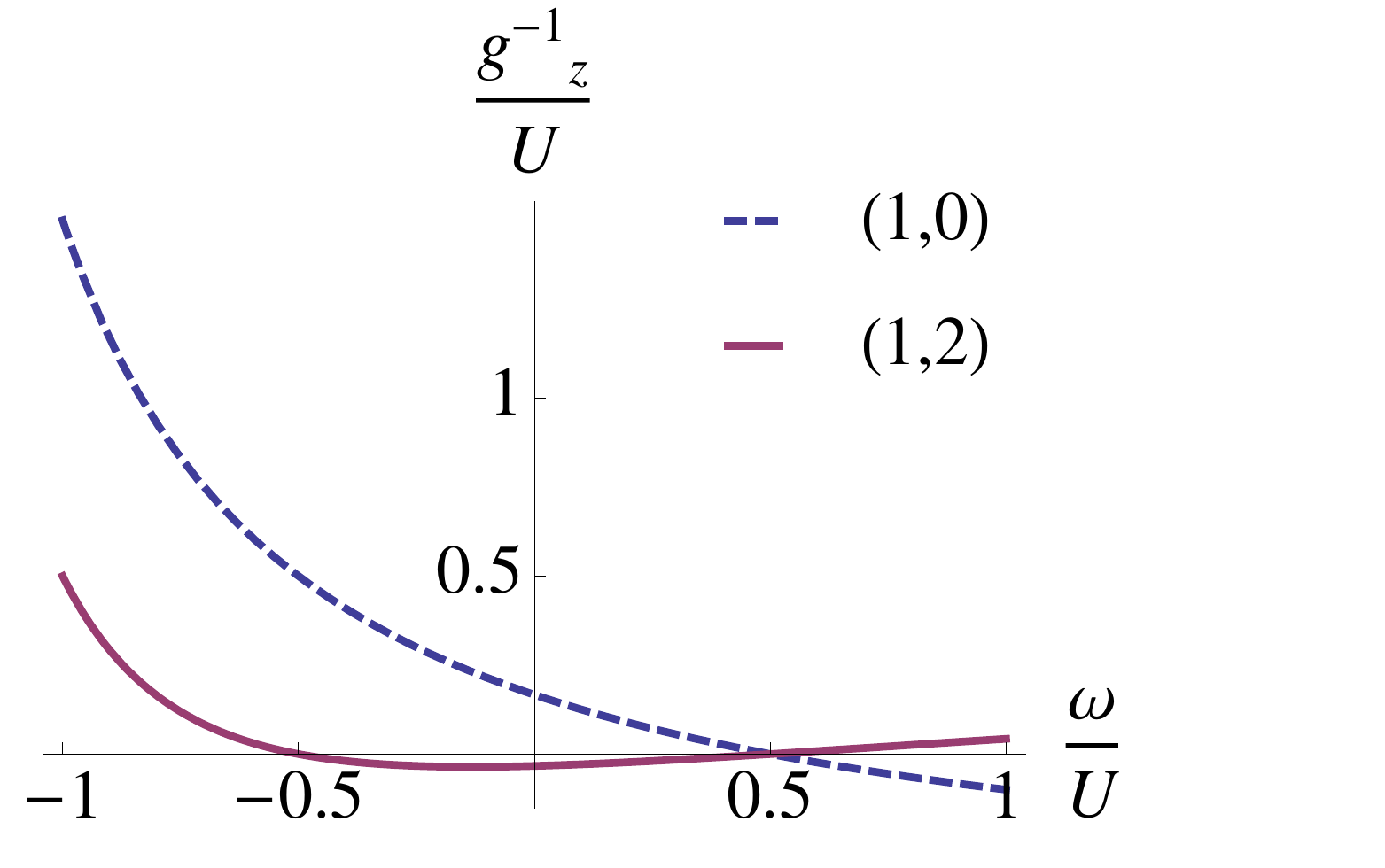}\\
(a)&(b)\\
\end{tabular}
\caption{(Color Online)  (a) The inverse onsite propagator $(g^{-1})_\al(\w)$ of band $\al$ for $N_\al=0,1,2$, and (b) the z component $(g^{-1})_z(\w)$, given in Eq.~\eqref{invg}.}
\label{invgfig}
\end{center}
\end{figure}

We consider equal intra--spin interaction, and parametrize the ratio of inter to intra-spin interaction as
\[U_{++}=U_{--}=U\,,\quad U_{+-}=\lm U\,,\]
then 
 \ben
 \lm N_{-\al}-1<{\mu_\al\over U}- N_\al<\lm N_{-\al}
 \,.
 \een
%then the excitation energies simply given by $ \xi_\al^{(\pm)}=\pm U/2$.
The interspin term in Eq.~\eqref{onsiteE} favors a ferromagnetic state with $N_+-N-$ maximized, but the intraspin term favors both $N_\pm$ to be minimized.  Below, we consider two limits. 

\subsubsection{Without interspin interaction\label{without}}
Taking the limit of Ref.~[\onlinecite{wongPRL13}],  we turn off the inter-spin interaction, setting  $\lm=0$, then the excitation energies are $\xi_\al(N_\al)=UN_\al-\mu_\al$ and the zeroth order ground state has filling factors $N_\al$ when $N_\al-1<{\mu_\al/ U}<N_\al$.  Setting ${\mu_\al}=U(N_\al-1/2)$, at the point where $\p E^{(0)}/\p N_\al=0$, the onsite propagators are given by
\begin{align}
g^{-1}_\al&=\frac{(\omega+1/2) (\omega-1/2)}{U(N_\al+1/2)+\omega}\,,\nn
\left\{\begin{array}{c}
  (g^{-1})_0 \\(g^{-1})_z
\end{array}\right\}
&=\frac{U  (4 \omega ^2-1)}{2 (1+2 N_++2 \omega ) (1+2 N_-+2 \omega )}\nn
&\left\{\begin{array}{c}
1+N_++N_-+2 \omega\\N_--N_+
\end{array}\right\}\,.
\label{invg}
\end{align}
and plotted these in Fig.~\ref{invgfig}. filling factors $(N,0)$, 
They show zeros at the quasiparticle and quasihole excitation energies except for the $(1,0)$ filling, where the zero for one hole band is missing in Fig.~\ref{invgfig}(b)(blue curve).  

\subsubsection{With interspin interaction}
Consider now the case  $0<\lm<1$ and $\mu_+=\mu_-$.
In this case it is useful to express the ground state energy in terms of $N=N_++N_-$, $N_z=N_+-N_-$,
\[
E^{(0)}\left(N,N_z\right)=N_z^2 \left(\frac{1- \lambda }{4}\right)+N^2 \left(\frac{1+ \lambda }{4}\right)-N \left(\frac{1}{2}+\mu \right)\,.
\]
For even $N$, it is energetically favorable to have $N_z=0$, then $\mu$ lies in 
\[N_+\lm-1<{\mu\over U}-N_+<N_+\lm\,,\quad N_+=N_-\,. \]
For $N$ odd, however, $N_z=0$ is not possible, and to minimize energy the ground states will have $N_+=N_-\pm1$, outside the regions given by the inequality above.  The phase diagram is given in Ref.~\onlinecite{isacssonPRB05}.

\begin{figure}[t]
\begin{center}
\begin{tabular}{cc}
\includegraphics[width=.5\linewidth]{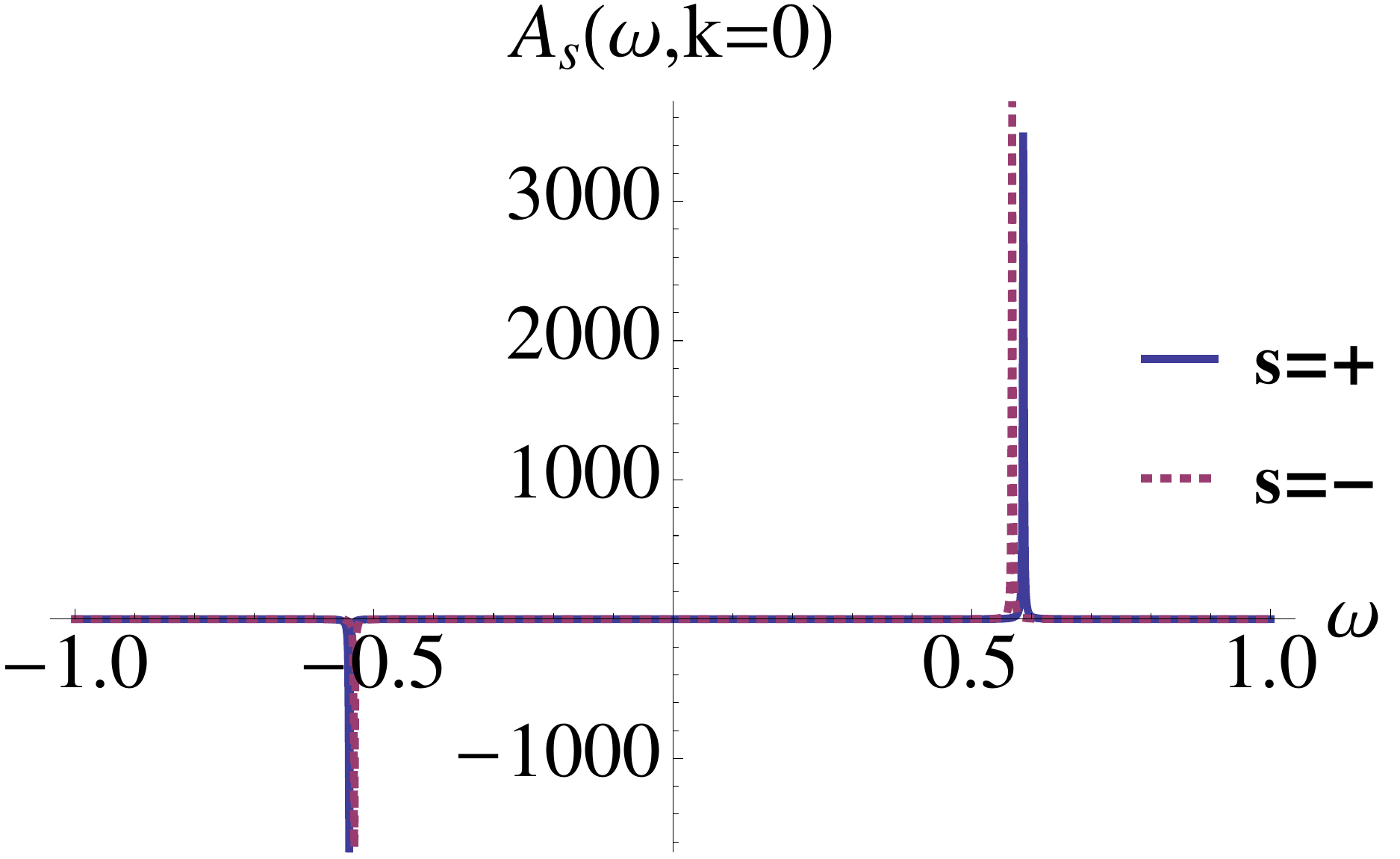}&
\includegraphics[width=.5\linewidth]{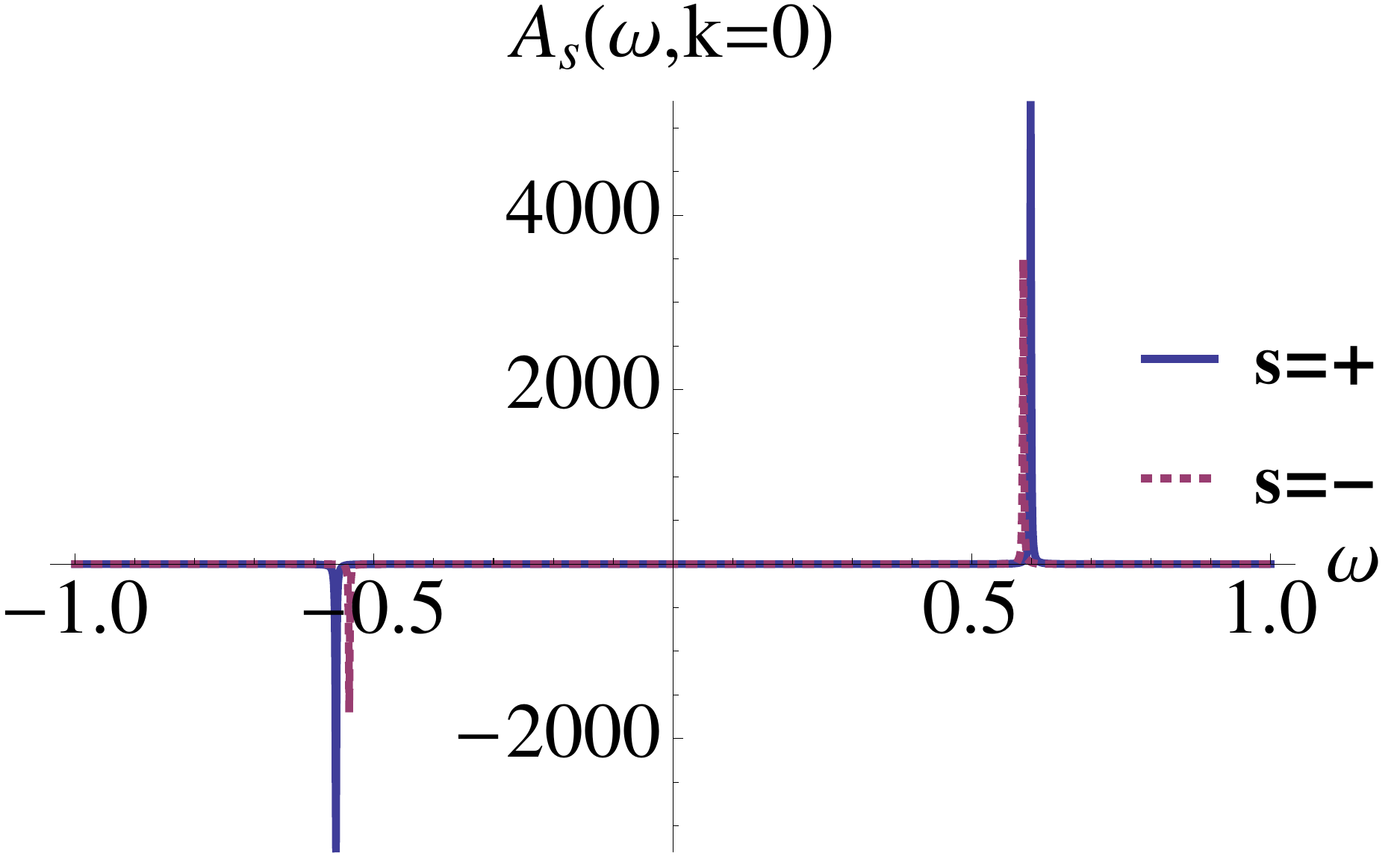}\\
(a)&(b)\\
\end{tabular}
\caption{(Color online) The spectral function ${A}_s(\w,\mbk=0)$ for $B=0, A=D=t$; $(\D/U,t/U)=.01$; (a) $(1,1)$ and (b) $(1,2)$ filling, with infinitesimal parameter $0^+=10^{-3}$.}
\label{spectralBHZ}
\end{center}
\end{figure}

Choosing a line $\mu(\lm)$ satisfying the above inequalities, the  particle and spin components of the inverse propagator are then given by 
\begin{align*}
(g^{-1})_0&=\mu(\lm)+\omega-\frac{U}{2}(N_++N_-)(\lambda +2)\nn
&+\frac{U}{2}\sum_\al\frac{N_\al (1+N_\al)}{1-N_{-\al}\lambda +(\mu(\lm) +\omega )/U}\,;\nn
(g^{-1})_z&=\frac{U}{2} (N_+-N_-) (\lambda -2)\nn
&+\frac{U}{2}\sum_\al\al\frac{N_\al (1+N_\al)}{1-N_{-\al}\lambda +(\mu(\lm) +\omega )/U} \,.
\end{align*}
So far, we have assumed a uniform ferromagnetic ground state in the $z$ direction defined in the hopping Hamiltonian.  Other types of magnetic order such as $XY$ ferromagnet and spin spirals are possible, but a rigorous computation of the phase diagram of the ground state magnetization is beyond the scope of this work.

\section{Results for $U_{+-}=0$}
\label{results}
%{without  inter-spin interaction}
\label{nointerspin}
Below we consider the hopping parameters $B=0$, $A=D=t$.  We first check that the Green function have only simple poles which results in sharp quasiparticle peaks in the spectral function. This is verified in   in Fig.~\ref{spectralBHZ}, where we  plot the band $s$ spectral functions at the origin of the BZ, $A_s(\w,\mbk=0)$, for the $(1,1)$ and $(1,2)$ filling.   Note that as the ratio $N_+/N_-$ decreases, so does the relative spectral weights, as expected, since $Z_{s-}\propto N_s$.

Next, we find the dispersions by solving Eq.~\eqref{disp} with the onsite propagators defined in Eq.~\eqref{invg}. First, we note that in the cases of equal filling fractions $(N_+,N_-)=(N,N)$, we can set $\mu_\al=\mu$. In this case, $(g^{-1}_z)=0$, so that there are no corrections to the spin texture, which is given by $\mbh_\mbk$, and we have
\[
(g^{-1})_0=-2UN_++\mu +\omega +\frac{N_+(N_++1)U^2}{U+\mu +\omega}\,.
 \]
The dispersions can be expressed for any $\mu$ as
\ben
{\e_\ksn+\mu\over U}=N_+-{1\over2}+{\bare_{\mbk s}\over2}+{n\over2}\sqrt{1-2 (1+2 N_+)\bare_{\mbk s}+\bare_{\mbk s}^2}\,,
\label{dispN}
\een 
where $\bare_{\mbk s}=\e^{(1)}_{\mbk s}/U$ are the noninteracting dispersions in units of $U$.  A plot of this dispersion as function of $\bare$ is given in Fig.~\ref{dispe}(a), and is very similar to the spinless case of Ref.~[\onlinecite{oostenPRA01}].\footnote{
Imaginary parts arise when the expression inside the square root becomes negative, indicating the superfluid transition.\cite{grassPRA11}  } 
\begin{figure}[t]
\begin{center}
\begin{tabular}{cc}
\includegraphics[width=.5\linewidth]{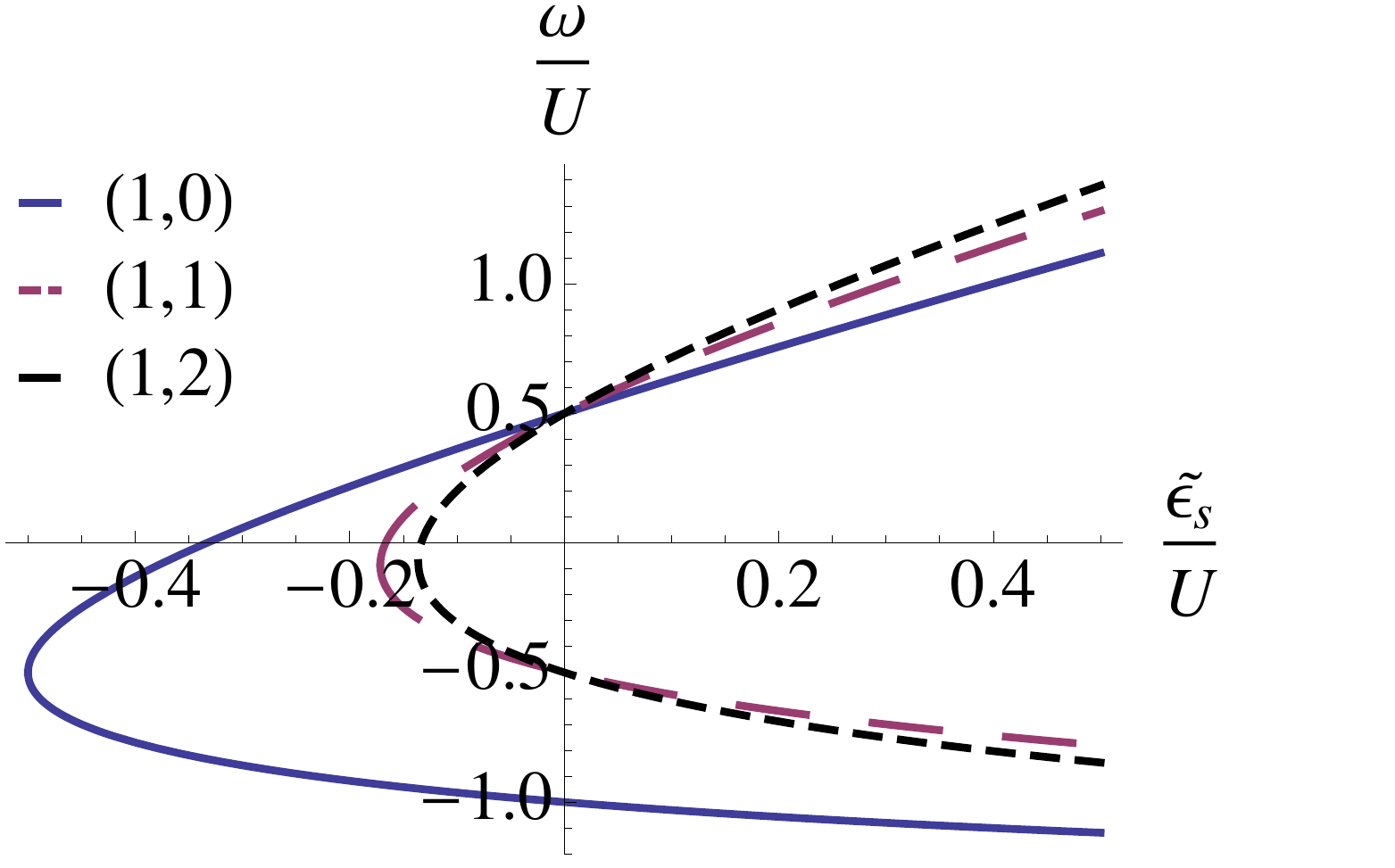}
&\includegraphics[width=.5\linewidth]{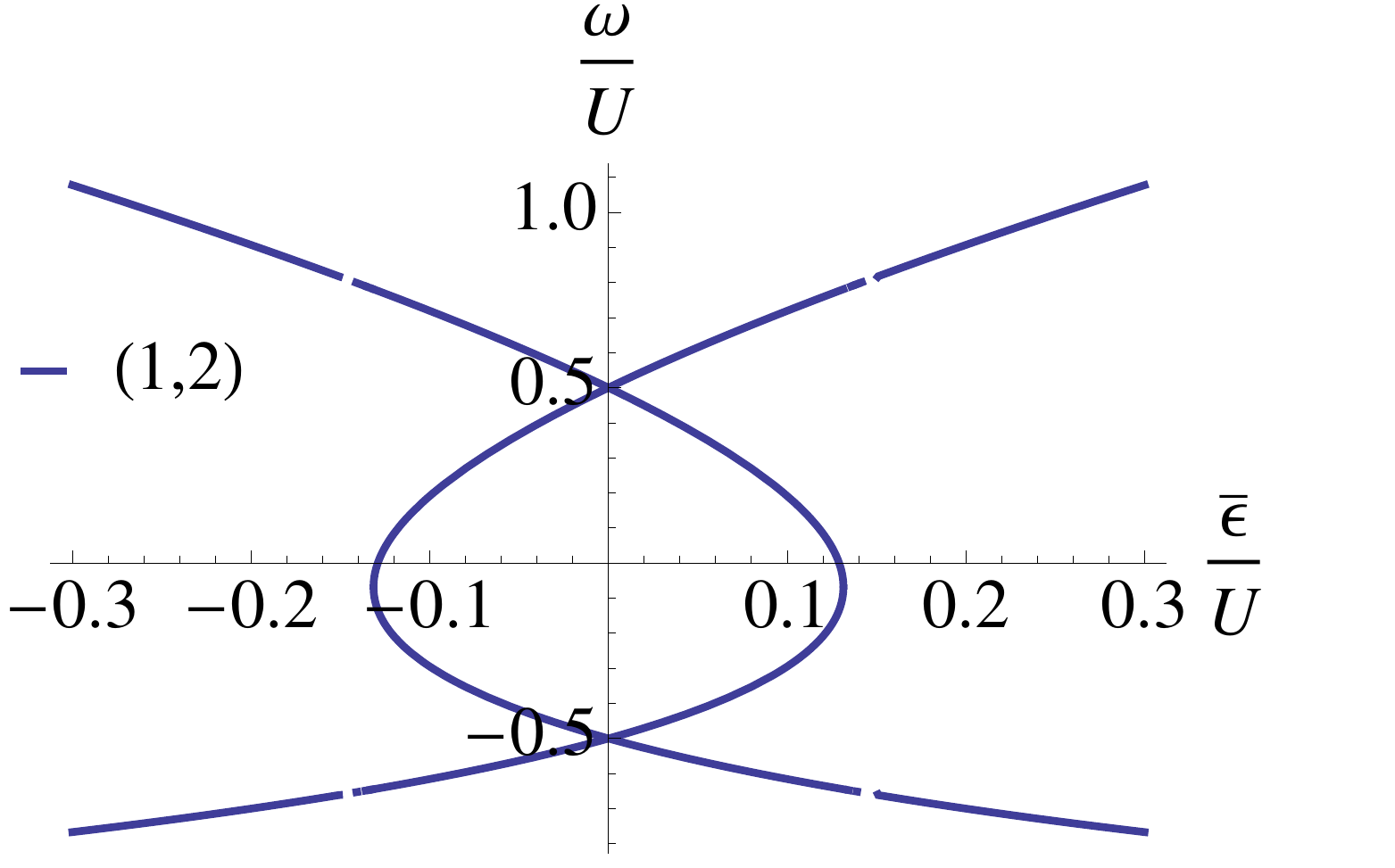}\\
(a)&(b)\\
\end{tabular}
\end{center}
\caption{(Color Online)
(a) A plot of the dispersions as function of $\til{\e}_s/U$ defined in Eq.~\eqref{eqdispe} for  $(N_+,N_-)=\{(1,0),(1,1),(1,2)\}$. 
(b) For a massless Dirac hopping Hamiltonian, the $(1,2)$ dispersion as function of $\bare/U$, given by the solution of Eq.~\eqref{diracdisp}}
\label{dispe}
\end{figure}
The simplest filling fraction which gives an out-of-plane spin texture is $(1,0)$, which represents the simplest ferromagnetic ground state.  This case was discussed in Ref.~[\onlinecite{wongPRL13}].  The quasiparticle dispersions can be expressed implicitly as 
\[
{\e_\ksn\over U}=\frac{1}{4} \left(-1+2 {\til{\e}_\ksn\over U}+n\sqrt{9+20{\til{\e}_\ksn\over U}+4\pfrac{ \til{\e}_\ksn}{U}^2}\right)\,,
\]
where $\til{\e}_\ksn$ is defined in Eq.~\eqref{eqdispe}, and we plot this function in Fig.~\ref{dispe}(a).  Note that there is a missing quasihole band, so that in this formula $(s,n)\neq(-,-)$, because there are no spin down atoms in the unperturbed $(1,0)$ ground state. However, the band which evolves from the spin down band in the lab frame is along $\mbf{f}_{\mbk,s=-,n}$ and may not correspond to  spin down in the lab frame.  As discussed in section\ref{fkz}, in this case, there is a uniform $f^z_\ksn$, so that the Chern numbers are zero.  

\subsection{Topologically trivial state:  (1,1) filling factor}
We first consider the $(1,1)$ filling factor at  $t/U=\D/U=.01$.  Although the Chern numbers are zero, there are still nontrivial, interaction induced effects.   First, we define the function $n_s(\w)=-Z_s(\w)\equiv-[\p_\w G_s^{-1}(\w)]^{-1}$,  given by
\[
 n_s(\w)={1\over\p_\w d(\w)}=-\frac{(3+2 \omega )^2}{1+12 \omega +4 \omega ^2}\,,
 \]
which  has no explicit $\mbk,s$ dependence, and can thus be interpreted as a distribution function in frequency space because the particle density on each hole band is given by $n_s(\e_{\mbk s-})$.   We plot this function in Fig~\ref{11density}(a) and note that it is positive at the hole bands energies, as it should be.  We plot the $\mbk$ space particle density distribution in Fig~\ref{11density}(b), which shows that the particle distribution is nearly uniform at $n_\mbk\approx2$, so that in terms of their $\mbk$ space occupation the quasiparticles have fermonic character.   However, the distribution is peaked at $(\pi,\pi)$, indicating a tendency to Bose condense at this point at superfluid transition.  

Computing the real space density numerically, we find for $t/U=.01$, that $N(\mu=U/2)=\int d \mbk/(2\pi)^2\,n_\mbk=2.03.$  Thus, there is a small correction to $\mu$ of order $t$ necessary to keep the density at integer value.  However, as shown in the plot  of $N(\mu)$ (computed numerically) in Fig.~\ref{chern}(d), only a small tuning of $\mu$ is needed to satisfy $N(\mu)=2$ to stay in the Mott insulating phase, which we will neglect for the purposes of this paper. 

As mentioned above, for this filling $\hat{\mbf{f}}_\ksn=\hat{\mbh}_\mbk\equiv\mbh_\mbk/|\mbh_\mbk|$, so that the Berry curvature is the same as the noninteracting case, which is zero for $B=0$.   The spin density is given by [cf.~Eq.~\eqref{spinxy}] $\mbs_\mbk=(n_{\mbk+}-n_{\mbk-})\hat{\mbh}_\mbk$, and is plotted in Fig~\ref{11density}(d).  The quantity $n_{\mbk+}-n_{\mbk-}$, can be thought of as a $\mbk$ space ``magnetization" and is plotted  in Fig~\ref{11density}(c).  Note that it is negative, so the spin density $\mbs_\mbk$ has the opposite orientation as $\mbh_\mbk$.  

\begin{figure*}[t]
\begin{center}
\begin{tabular}{cc}
\includegraphics[width=.4\linewidth]{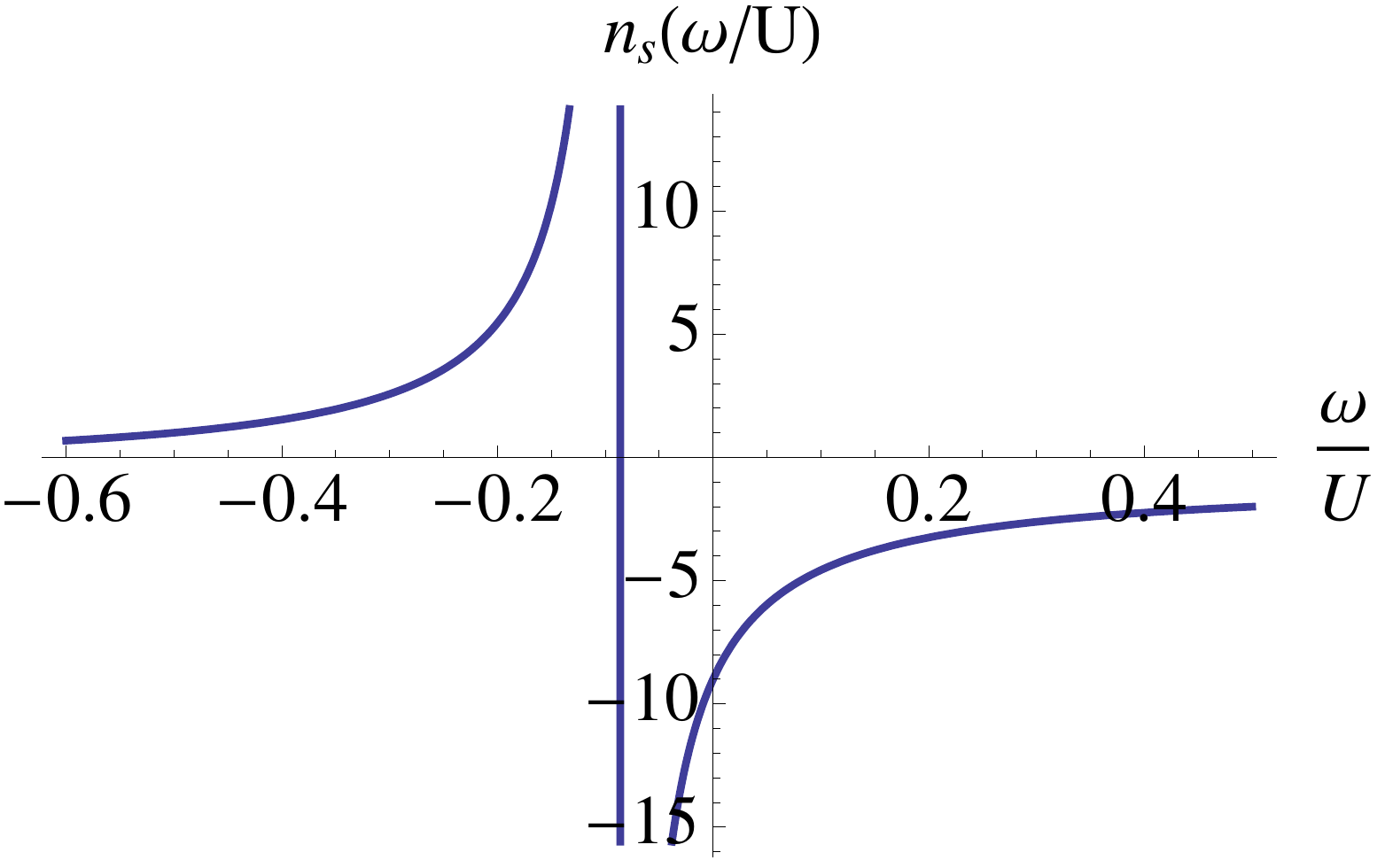}
&\includegraphics[width=.4\linewidth]{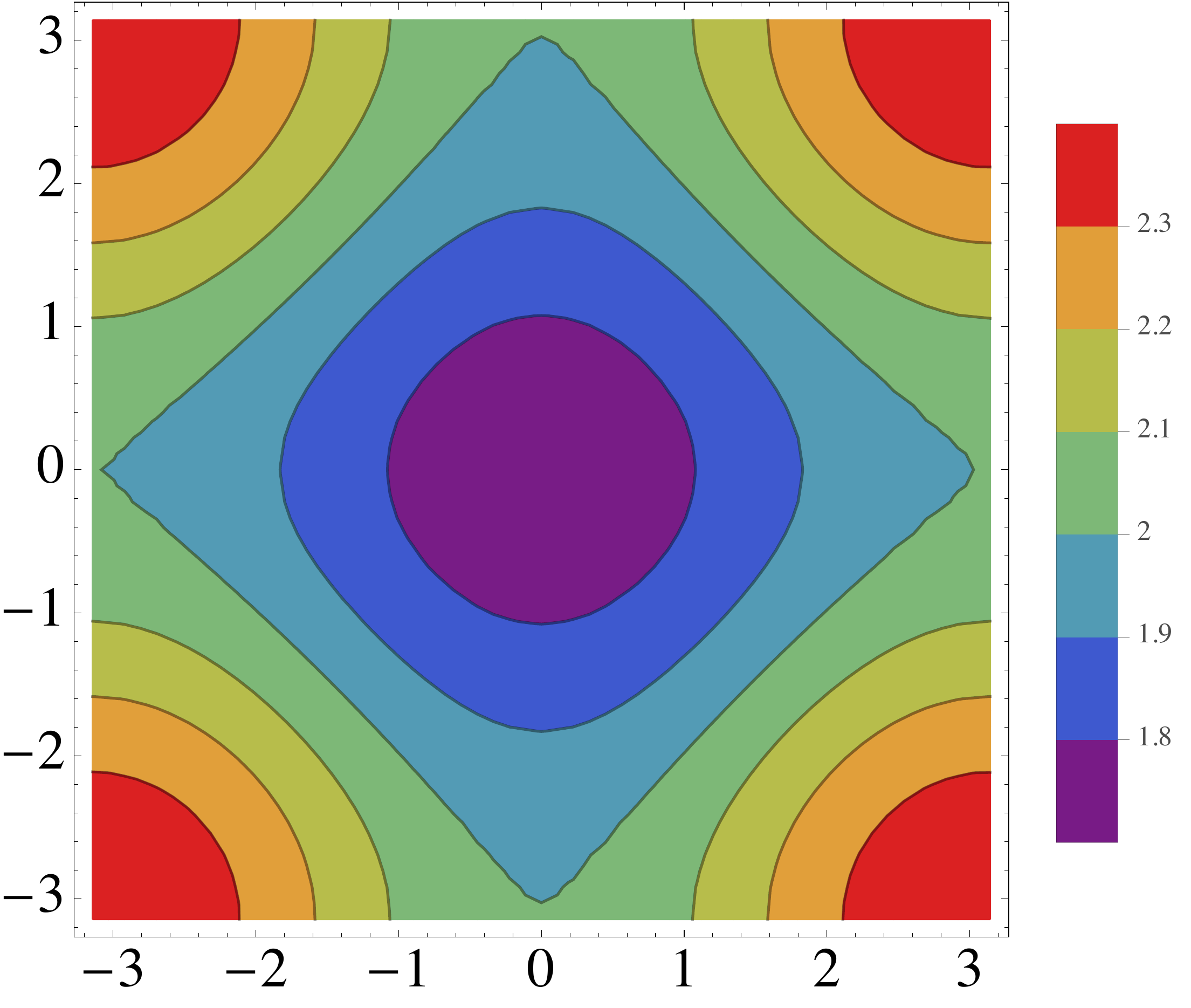}\\
(a)&(b)\\
\includegraphics[width=.4\linewidth]{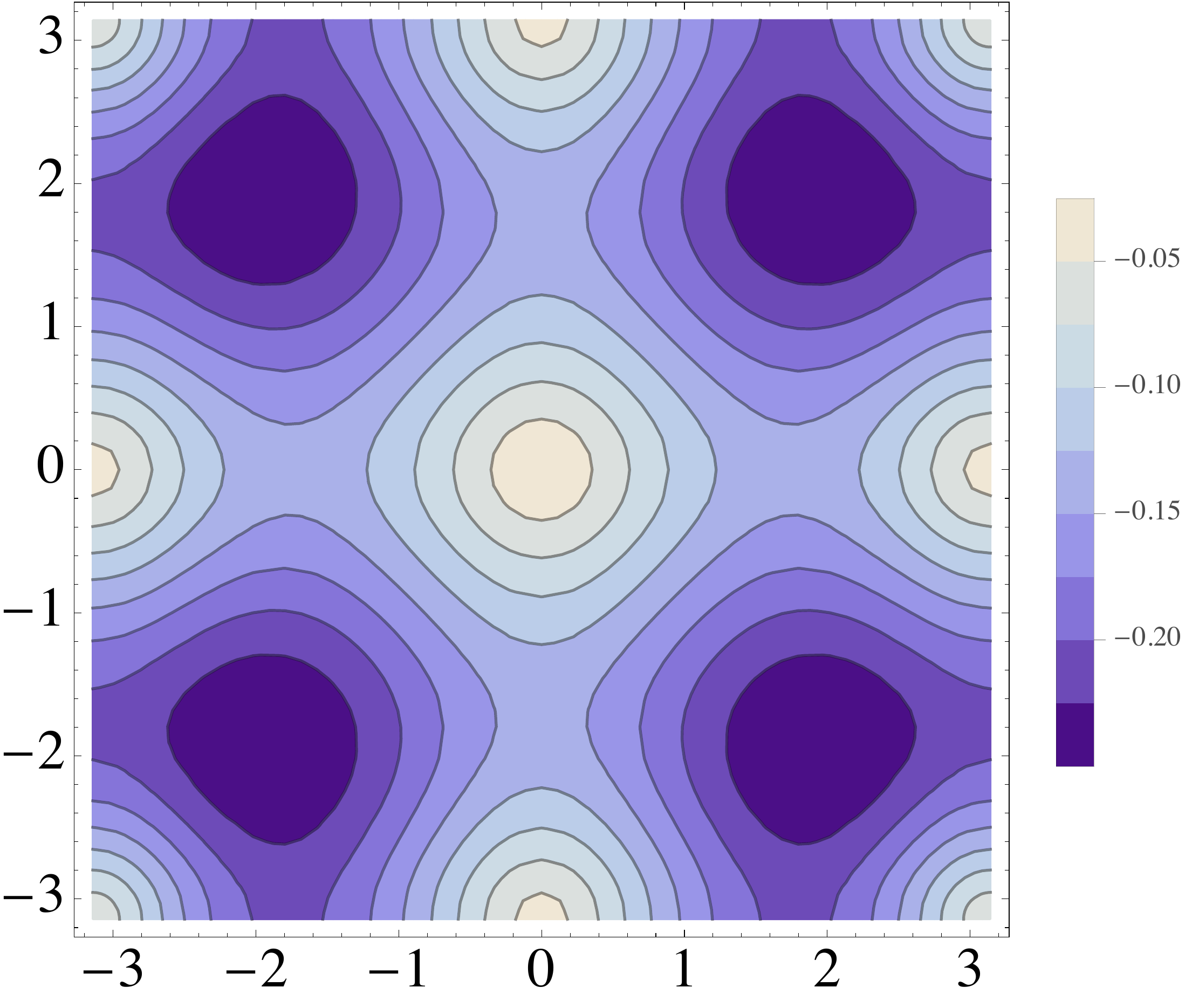}&
\includegraphics[width=.4\linewidth]{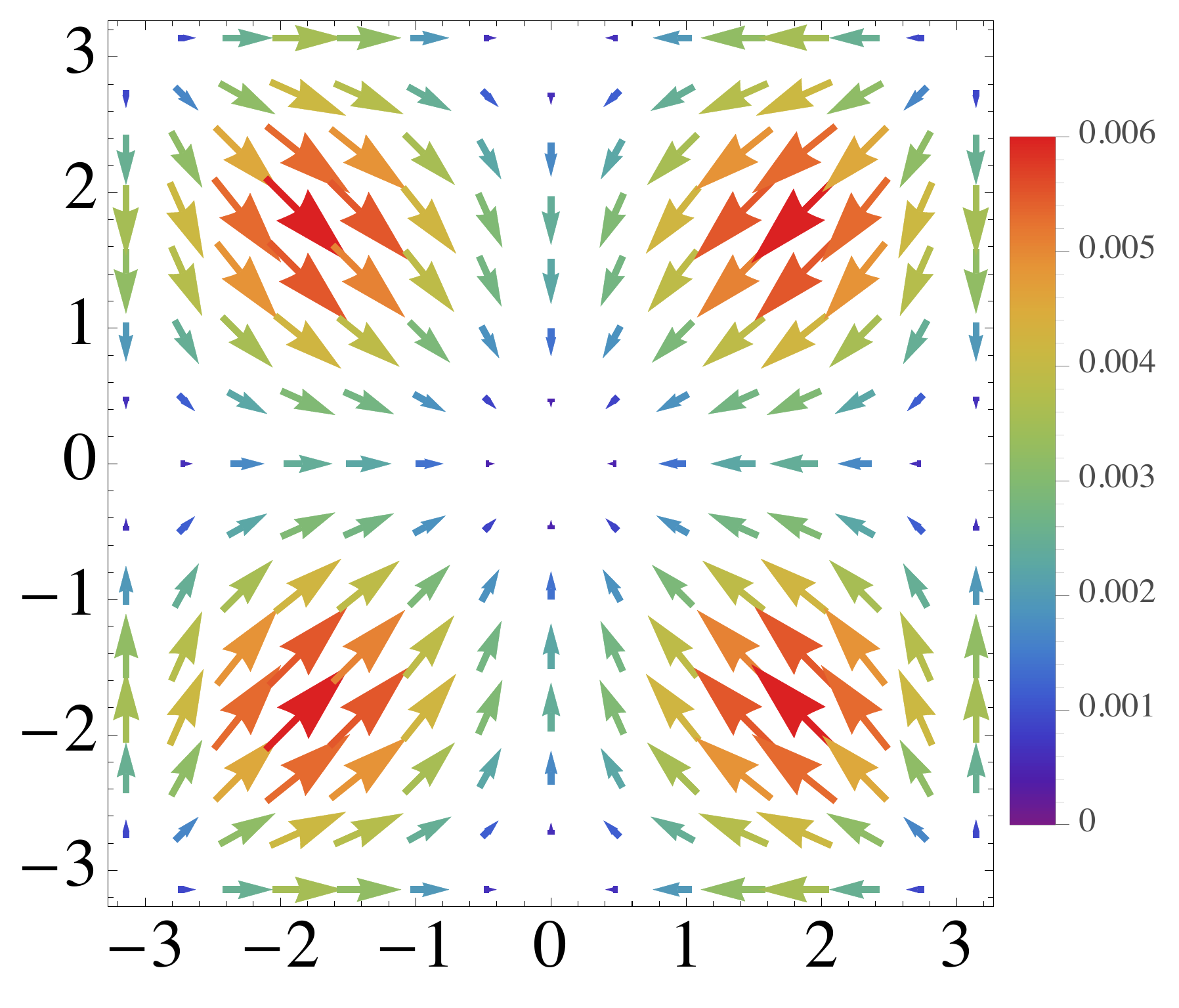}\\
(c)&(d)
\end{tabular}
\caption{(Color Online) For the filling fraction $(1,1)$ and $t/U=\D/U=.01$:  (a) Quasimomentum distribution of particle density  $n_\mbk$. (b) The difference in band $s$ densities $n_{\mbk+}-n_{\mbk-}$ (c) Quasimomentum distribution of spin density (in units of $\hbar$) $\mbf{s}_\mbk$.  (d) The in-plane spin density, $s_{x,y}(\mbk)$, where the color and size represent the in-plane magnitude (in units of $\hbar$).  In plots (b--d), the axes are wave vectors in the Brillouin zone in units of inverse lattice spacing.}
\label{11density}
\end{center}
\end{figure*}

\begin{figure*}[t]
\begin{center}
\begin{tabular}{cc}
\includegraphics[width=.4\linewidth]{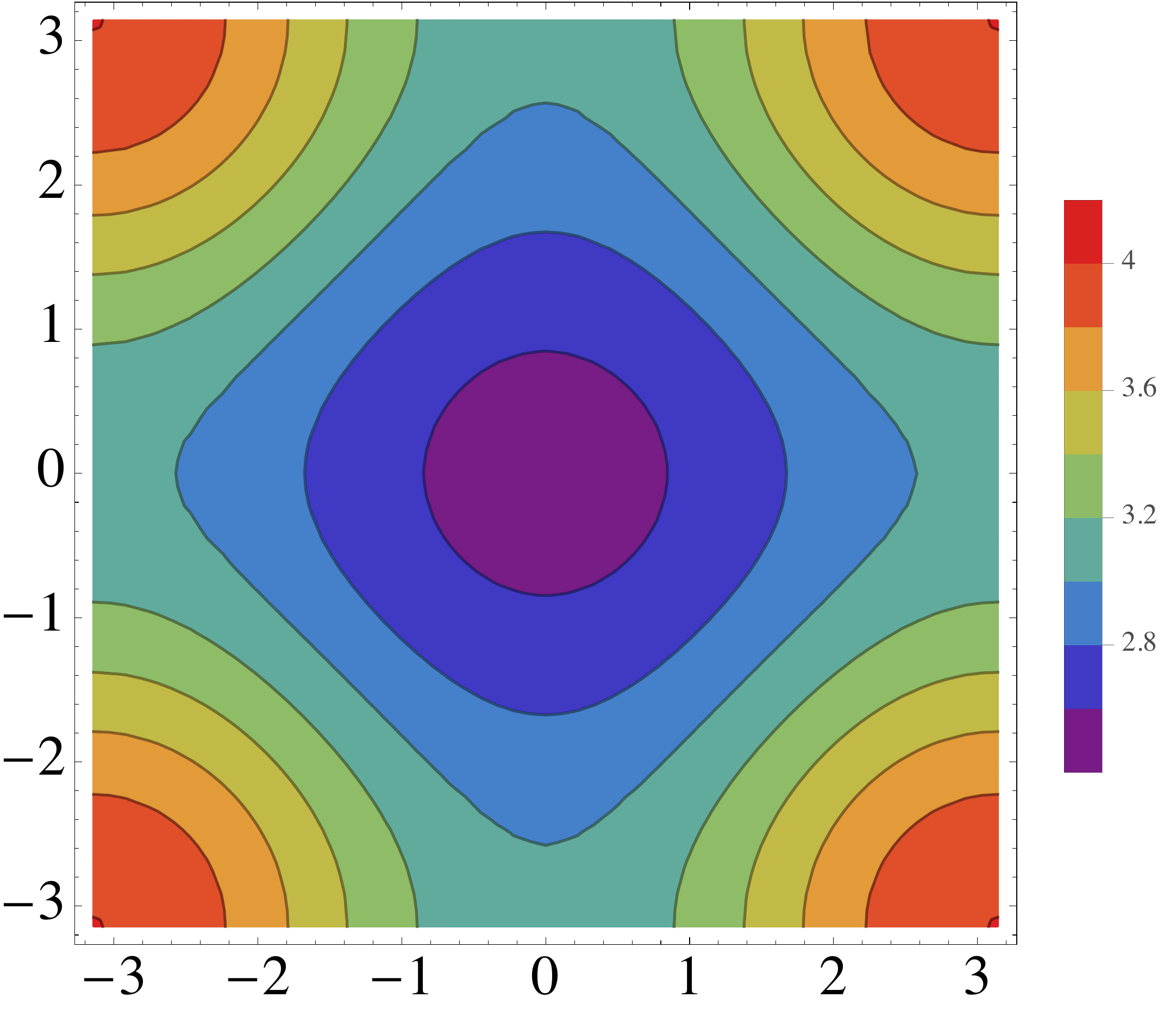}
&\includegraphics[width=.4\linewidth]{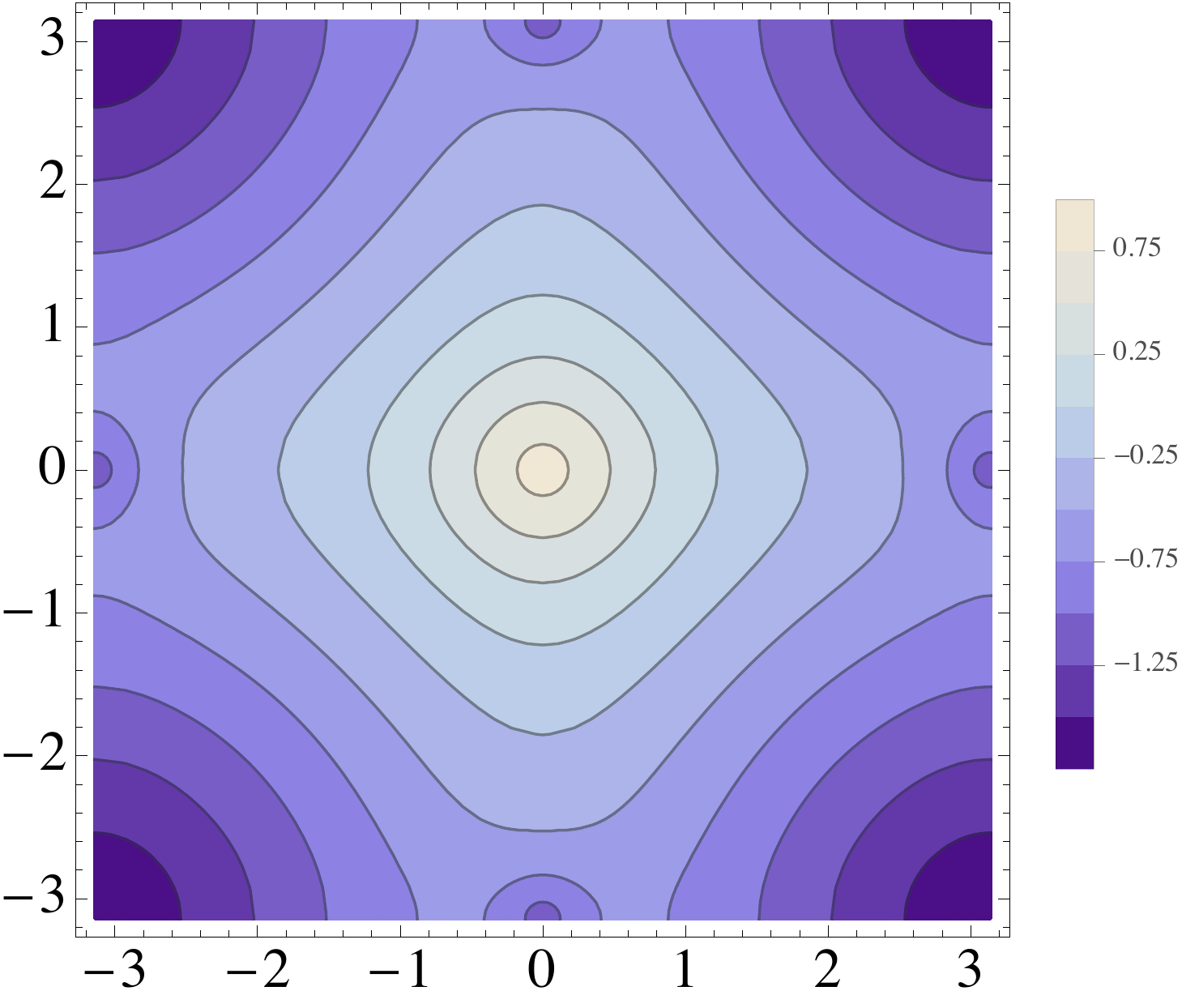}\\
(a)&(b)\\
\includegraphics[width=.4\linewidth]{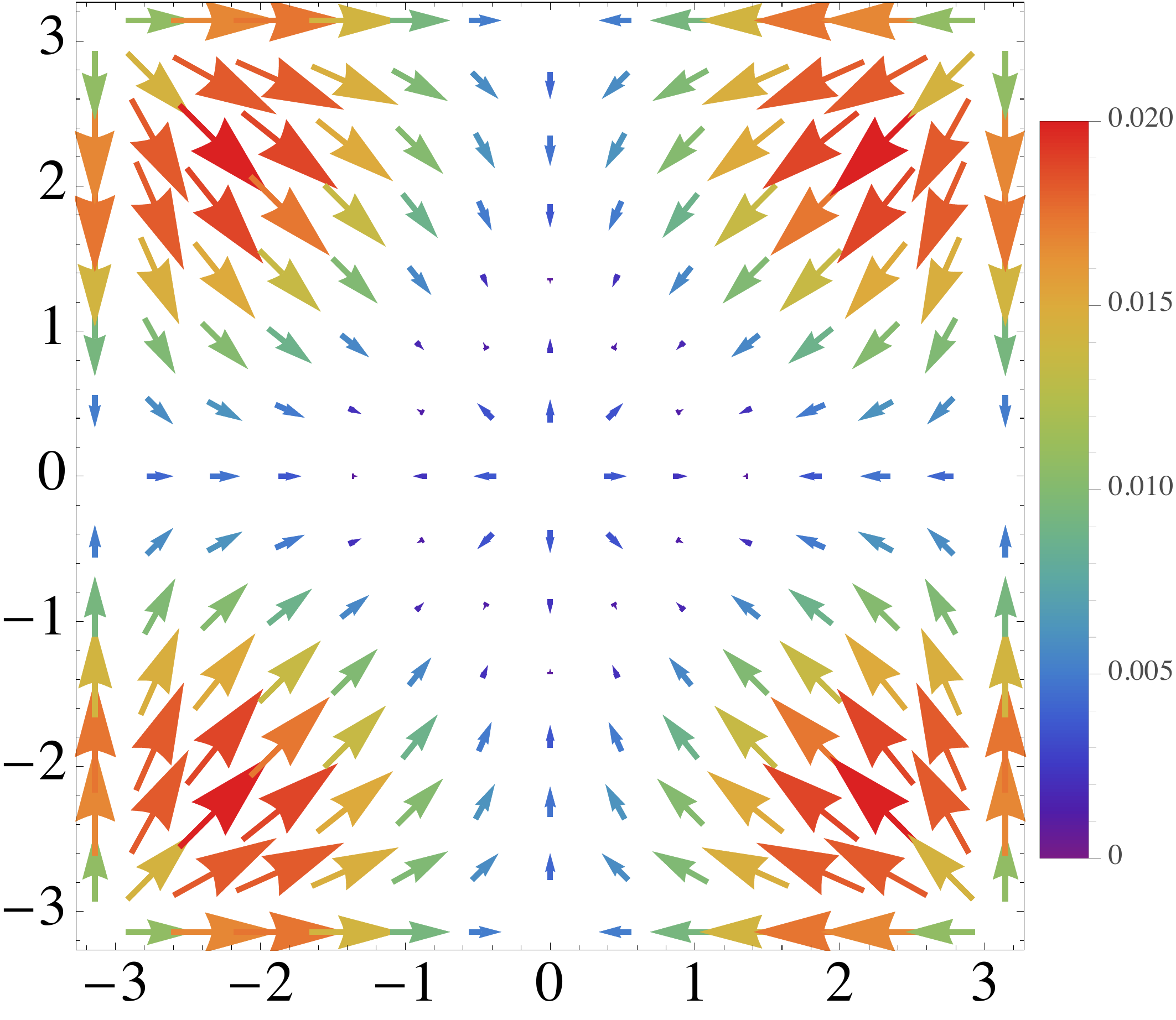}&
\includegraphics[width=.4\linewidth]{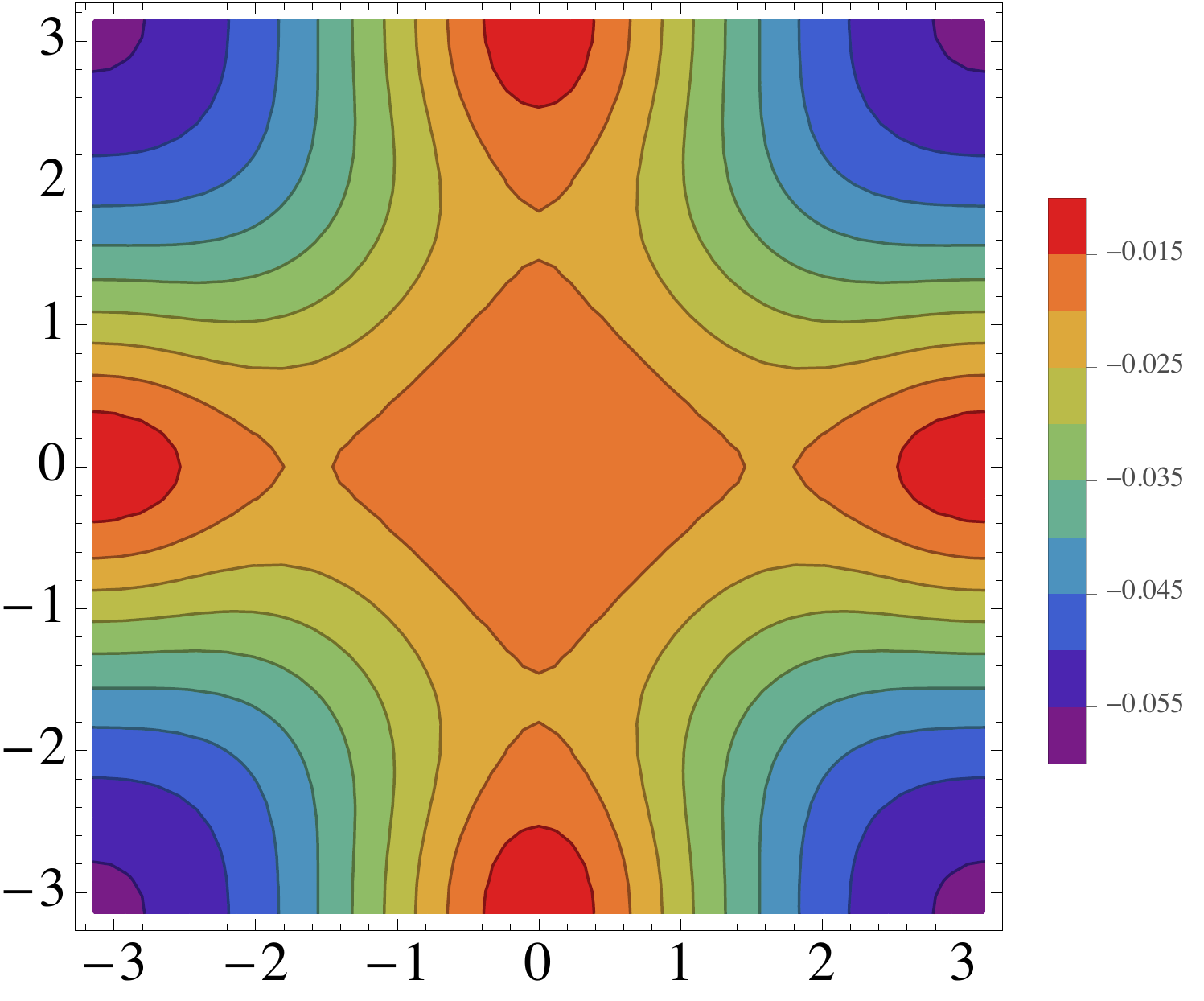}
\\
(c)&(d)
\end{tabular}
\caption{(Color Online) For the filling fraction $(1,2)$ and $t/U=\D/U=.01$: (a) Quasimomentum distribution of the $(1,2)$ particle density distribution $n_\mbk$ and (b) the difference in band $s$ densities $n_{\mbk+}-n_{\mbk-}$; Quasimomentum distribution of (c) in-plane ${s}_{x,y}(\mbk)$, where the color and size represent the in-plane magnitude(in units of $\hbar$).  The vector field changes from pointing outward in the center at $\mbk=(0,0)$ to inward away from the center.  (d) Out-of-plane spin density ${s}_{z}(\mbk)$ (in units of $\hbar$).  In plots (b--d), the axes are in wave vectors in the Brillouin zone in units of inverse lattice spacing.}
\label{12density}
\end{center}
\end{figure*}

\begin{figure*}[t]
\begin{center}
\begin{tabular}{cc}
\includegraphics[width=.4\linewidth]{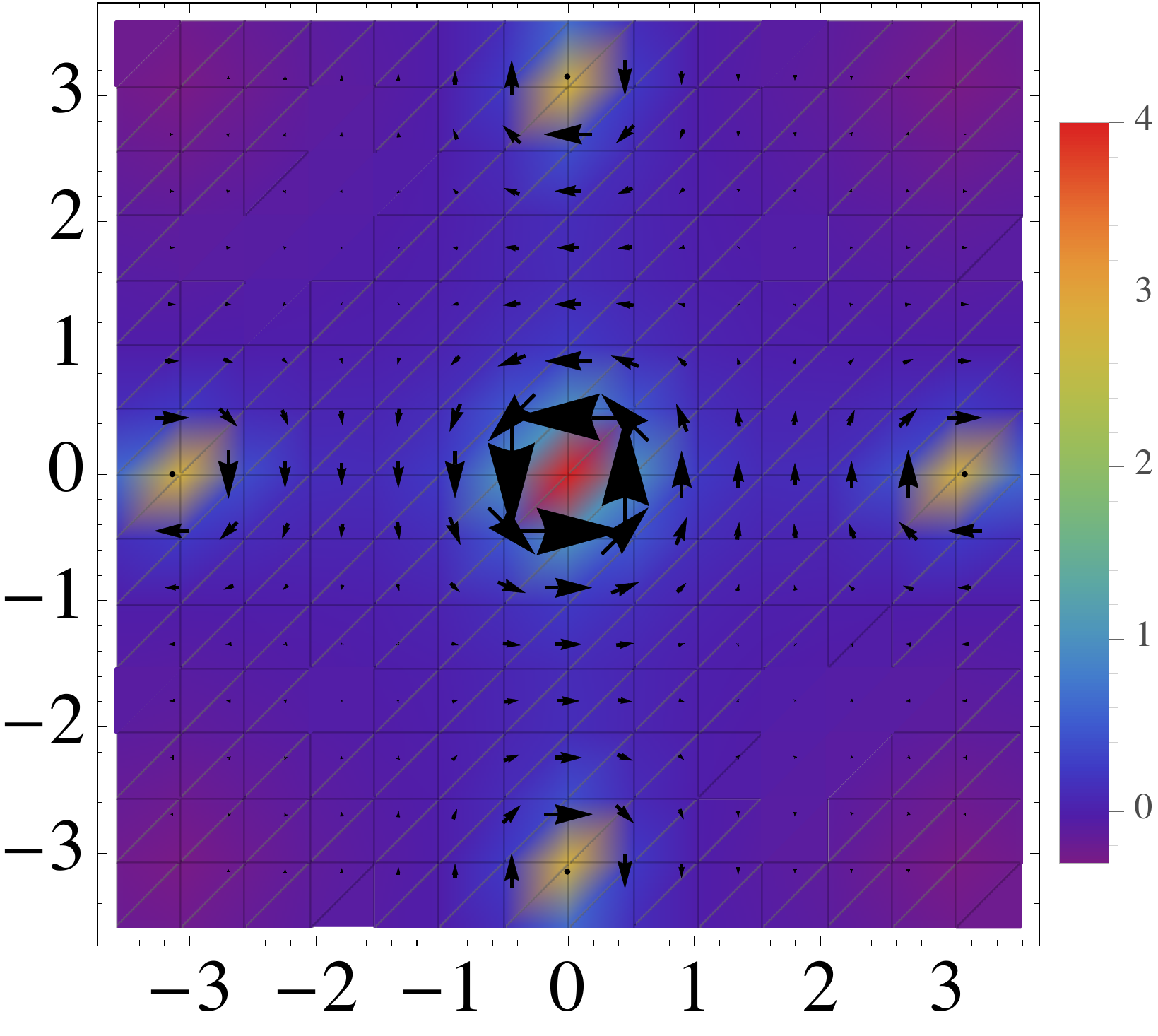}
&\includegraphics[width=.4\linewidth]{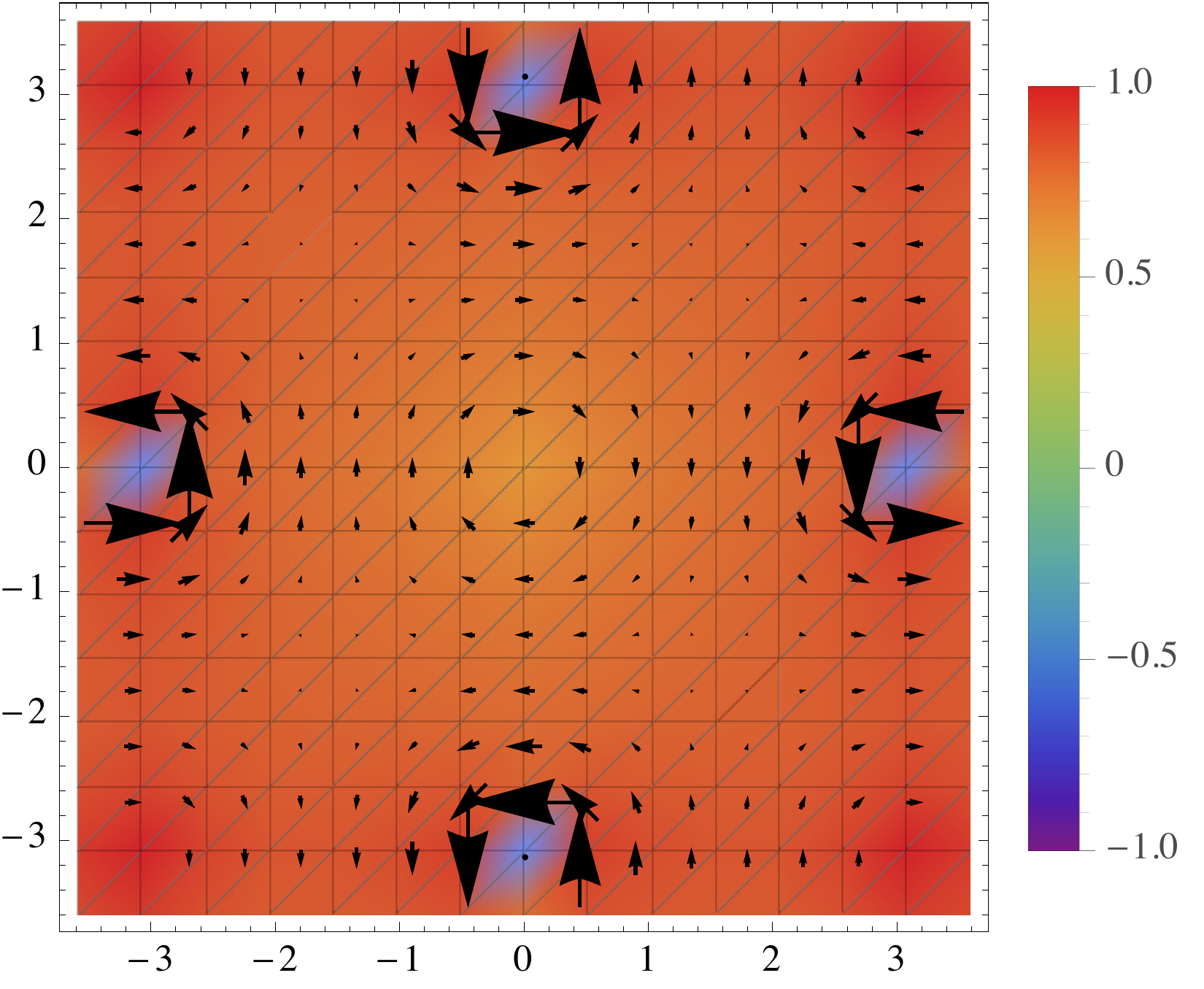}\\
(a)&(b)\\
\includegraphics[width=.4\linewidth]{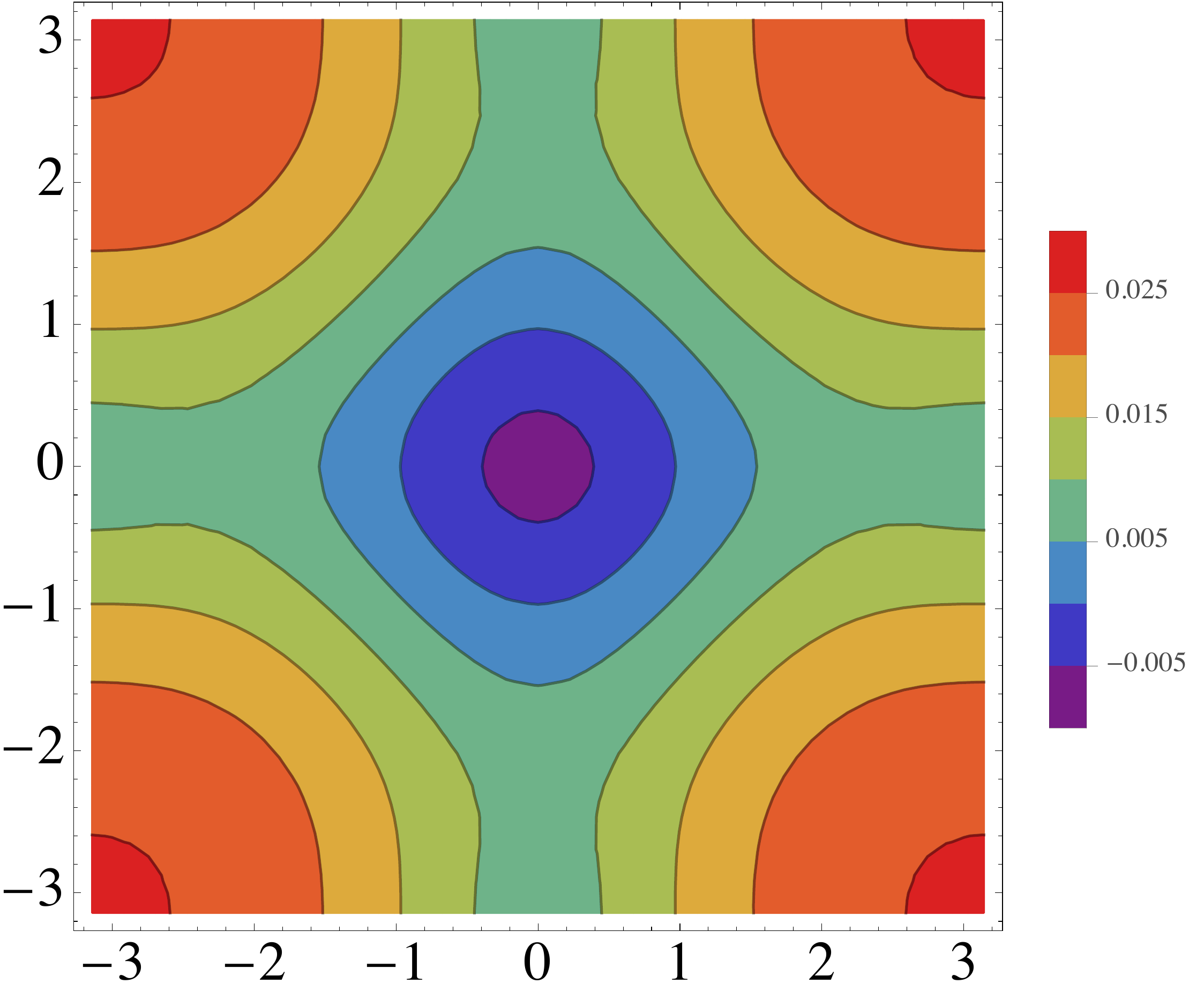}&
\includegraphics[width=.4\linewidth]{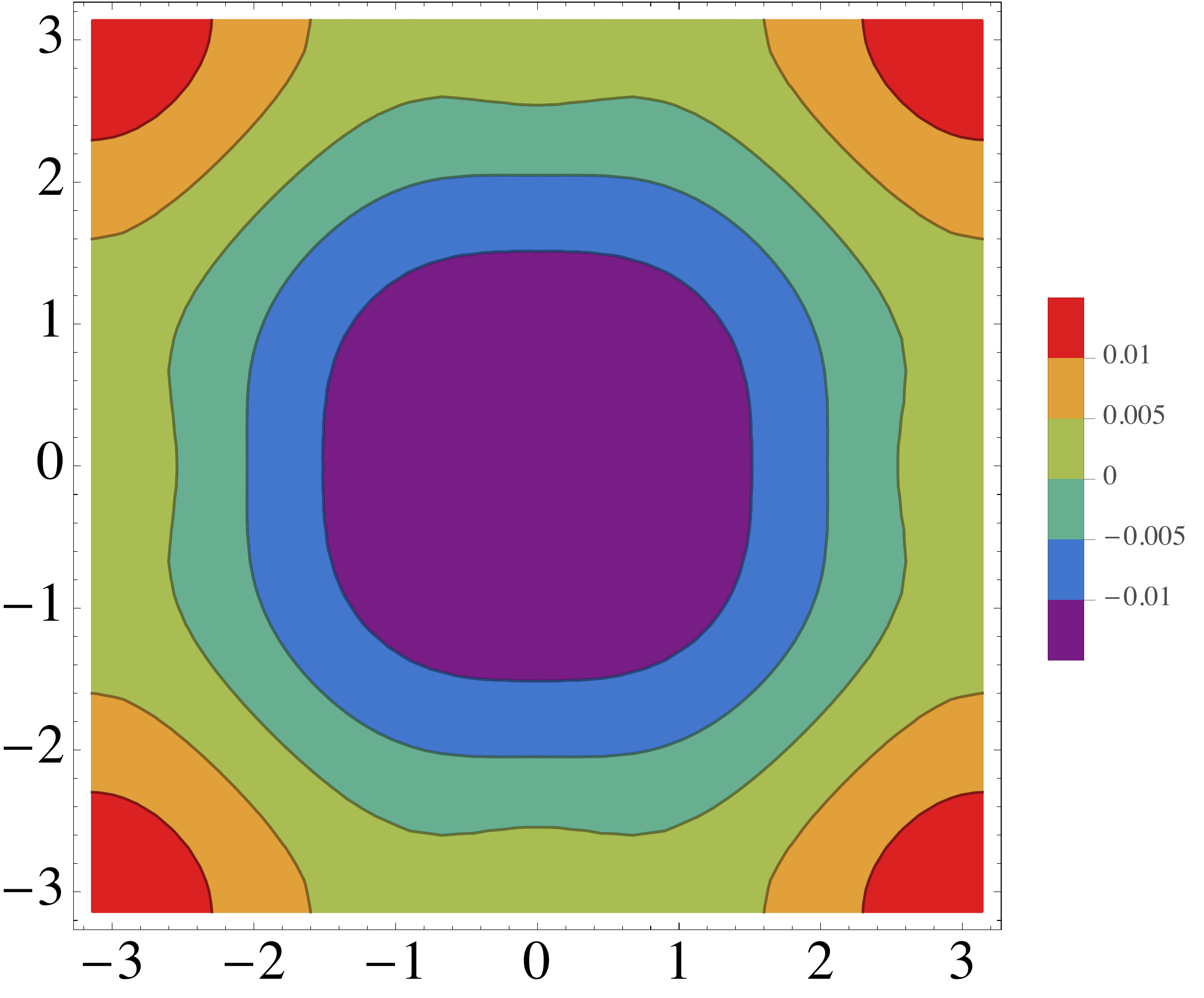}\\
(c)&(d)
\end{tabular}
\caption{(Color Online) Top row: The $(1,2)$ interacting Berry electromagnetic fields on the hole bands, in units of lattice spacing squared, for (a) $s=-$ and (b)$s=+$. The vector field is the Berry electric field $\mcal{E}_\ksn$ and the color density plot is the total Berry curvature $\mcal{C}_\ksn$. Bottom row: {The SO gaps $f^z_\ksn/U$ of the quasiparticles in the Mott insulator, including the corrections to the hopping textures due to interactions, for the filling fraction $(1,2)$ and $D/U=\D/U=.01$:  (c) $f^z_{\mbk s-}/U$ and   (d) $f^z_{\mbk s+}/U$}. In plots (b--d), the axes are wave vectors in the Brillouin zone in units of inverse lattice spacing}
\label{berry}
\end{center}
\end{figure*}

\subsection{Topological state:  (1,2) filling factor\label{12filling}}

Next, we consider the $(1,2)$ case, which will be representative of a generic ferromagnetic Mott insulating state with filling factors $N_+\neq N_-$.  This state has purely interaction-generated SO gaps $f^z_{\mbk s-}$, plotted in Fig.~\ref{berry}.   The dispersions are plotted in Fig.~\ref{SOgap}, where it is seen that the hole bands actually have positive curvature.  We plot the density, $\mbk$ space magnetization, in--plane and out--of--plane spin density in Fig.~\ref{12density}(a--d).  In this case, the $\mbk$ space magnetization changes sign, causing a line of defects where the in--plane spin density vanishes, while the  the out--of--plane spin density points everywhere down along the  $z$ axis.   

We plot the Berry electric field $\mcal{E}_\ksn$ and the total Berry curvature $\mcal{C}_\ksn$ in Fig.~\ref{berry}. As in the noninteracting case,  they are concentrated at the high symmetry points $\mbk_i$.  The electric field also form vortices around these points, which is consistent with the frequency--momentum space Faraday's law\cite{shindouPRL06}
\[
[\p_\w\mcal{B}_{\mbk s}^z(\w)+\bsdel_\mbk\times\mcal{E}_{\mbk s}(\w)]\Big|_{\w=\e_\ksn}=0\,,
\]
which follows by virtue of their definition in terms of gauge fields.

Next we consider the interacting Chern numbers, computed similarly to Table \ref{topcharge}(right)  by summing the topological charge of each vortex,
\[
C_\al={\al\over2}\left[p_\al(0,0)+p_\al(\pi,\pi)-p_\al(0,\pi)-p_\al(\pi,0)\right]\,,
\]  
where the gaps are given by $f^z_\al(\mbk_i)=\D/2-(g^{-1})_z[\e_\al(\mbk_i)]$, and the energies at the gap locations are given by solving  $\e^{(1)}_\al(\mbk_i)-g_\al^{-1}(\w)=0$.    The analytical expression for the SO gap at $(0,0)$ is given by 
\begin{align}
f^z_{\alpha }(0,0)&=\frac{\Delta }{2}-\sum_{\al'}\al'\frac{F_{\alpha }(D,\Delta )^2/4-1}{2+4 N_{\al'}+F_{\alpha }(D,\Delta )}\,;\nn
F_{\alpha }(D,\Delta )&=8 D+\alpha  \Delta \nn
&-\sqrt{(2+8 D+\alpha  \Delta )^2+8 (8 D+\alpha  \Delta ) N_{\alpha }}\,.
\label{fz5}
\end{align}
On each SO band, the transition to the topological phase with finite $C_\al(D,\D)$ occurs in a similar manner to the noninteracting case, when the topological charges from the $(0,0)$ and $(\pi,\pi)$ vortices cancel while the other contributions add; specifically, $C_\al(D,\D)=0$ when $p_\al(0,0)=p_\al(\pi,\pi)$, and $C_\al(D,\D)=\pm1$ when $p_\al(0,0)=-p_\al(\pi,\pi)$, while $p_\al(0,\pi)=p_\al(\pi,0)=$sgn$\D$.  This scenario is verified by the plots of $f^z_\al(\mbk_i)$ in Fig.~\ref{chern}(a,b).  For fixed $\D$, this transition occurs at a finite value of $D$ that is needed to overcome a uniform SO gap imposed by $\D$ everywhere.   

The spin Chern number $C_z=C_+-C_-$ is  plotted as a function of $D$ and $\D$ in Fig.~\ref{Cz12}, which shows that $C_z=2$sgn$\D$ in a large part of the  parameter space, consistent with the discussion above.  {The plot terminates where the SO gap and hence the dispersions [cf.~Eq.~\eqref{eqdispe}] acquires imaginary parts, indicating the superfluid transition.} We have thus shown that a topological phase transition to a ground state with nonzero spin Chern number occurs in the ferromagnetic Mott insulating phase, which we again emphasize is a purely interaction effect.  

On the other hand, the total Chern number $C_0=C_++C_-$ vanishes in the whole parameter space.  This can be understood from Eq.~\eqref{fz4}, which shows that to leading order in the hopping parameters $D/U$ and $\D/U$, $f^z_\al$ is independent of $\al$, so that $C_+=-C_-$ [cf.~Eq.\eqref{C}].  Higher order corrections to Eq.~\eqref{fz4} introduces a small dependence on $\al$, as shown in Eq.~\eqref{fz5}, which has the leading order expansion in $D$ or $\D$ given by 
 \begin{align}
f^z_{\alpha }(0,0)&=\frac{4 D \left(N_+- N_-\right)+\left(N_++ N_-\right)\Delta /2 }{2 N_{-\alpha }}\nn
&+2 D \alpha  \Delta \left(\frac{ N_--N_+ }{N_{-\alpha } }+\frac{N_+{}^2-N_-{}^2\text{  }}{N_{-\alpha }^2 }\right)\nn
&+O\left(D^2\right)+O\left(\Delta ^2\right)\,.
\end{align}
The first term agrees with Eq.~\eqref{fz4}, while the second shows an $\al$ dependence.   However, the point where the topological transition occurs, where $f^z_\al=0$, remains the same on each SO band, as shown in Fig.~\ref{chern}(c).  The fact that we do not find the case with an odd Chern number, e.g. , $C_+=1$ and $C_-=0$, is consistent with a general argument that the integer Hall conductivity (given here by $C_0$) for bosons must be even integer valued.\cite{senthilPRL13}
\begin{figure}[t]
\begin{center}
\begin{tabular}{cc}
\includegraphics[width=.5\linewidth]{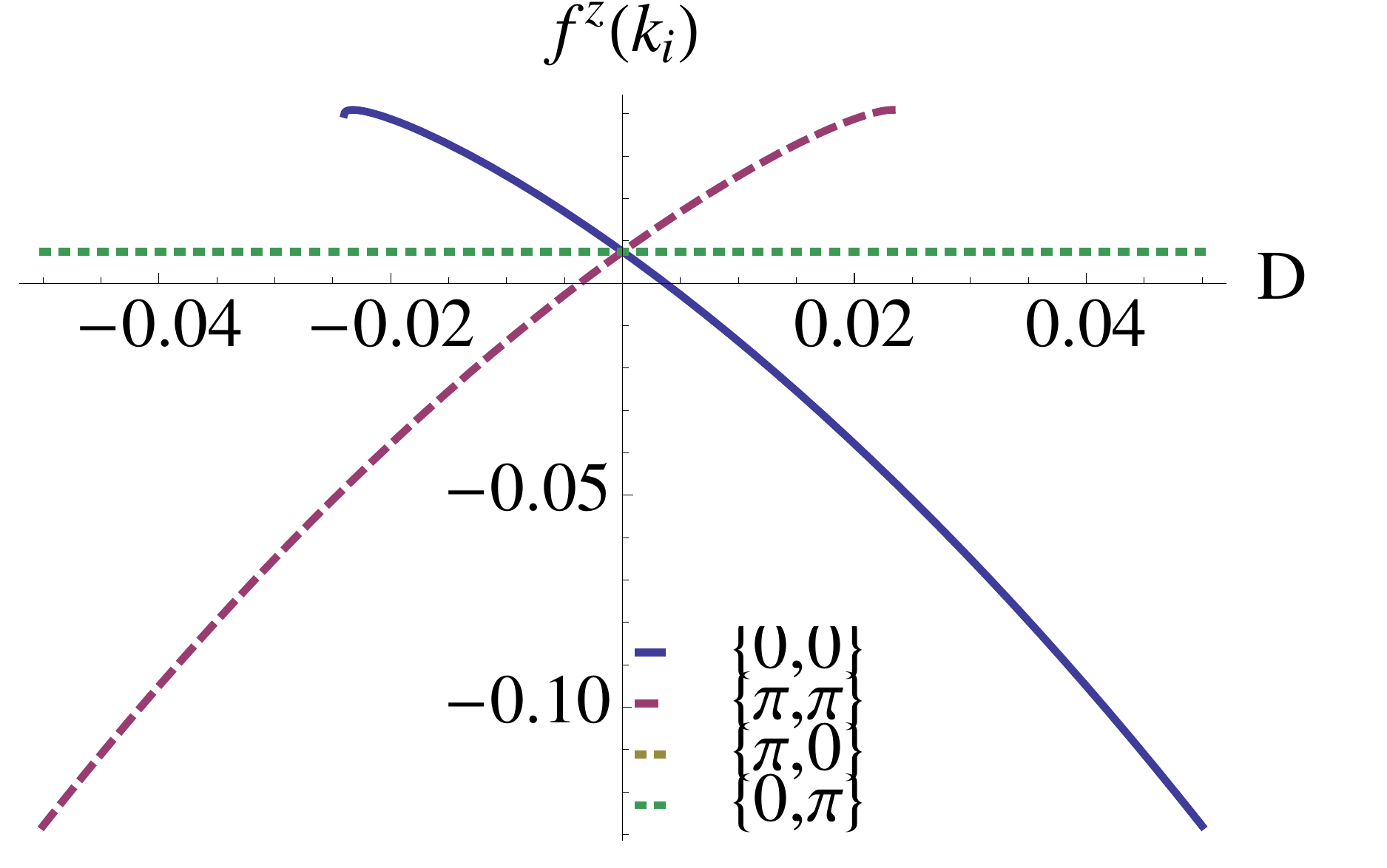}&
\includegraphics[width=.5\linewidth]{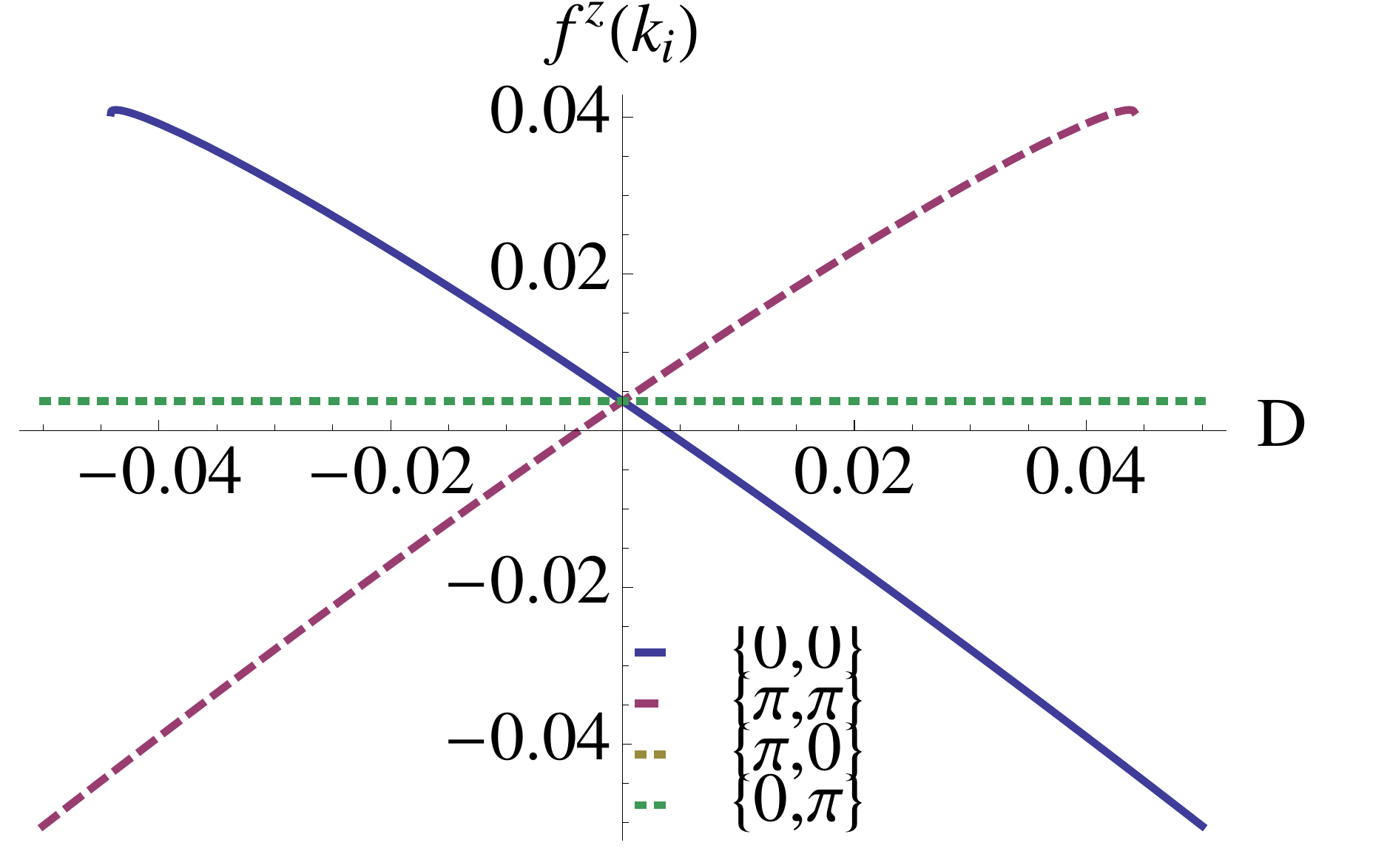}\\
(a)&(b)\\
\includegraphics[width=.5\linewidth]{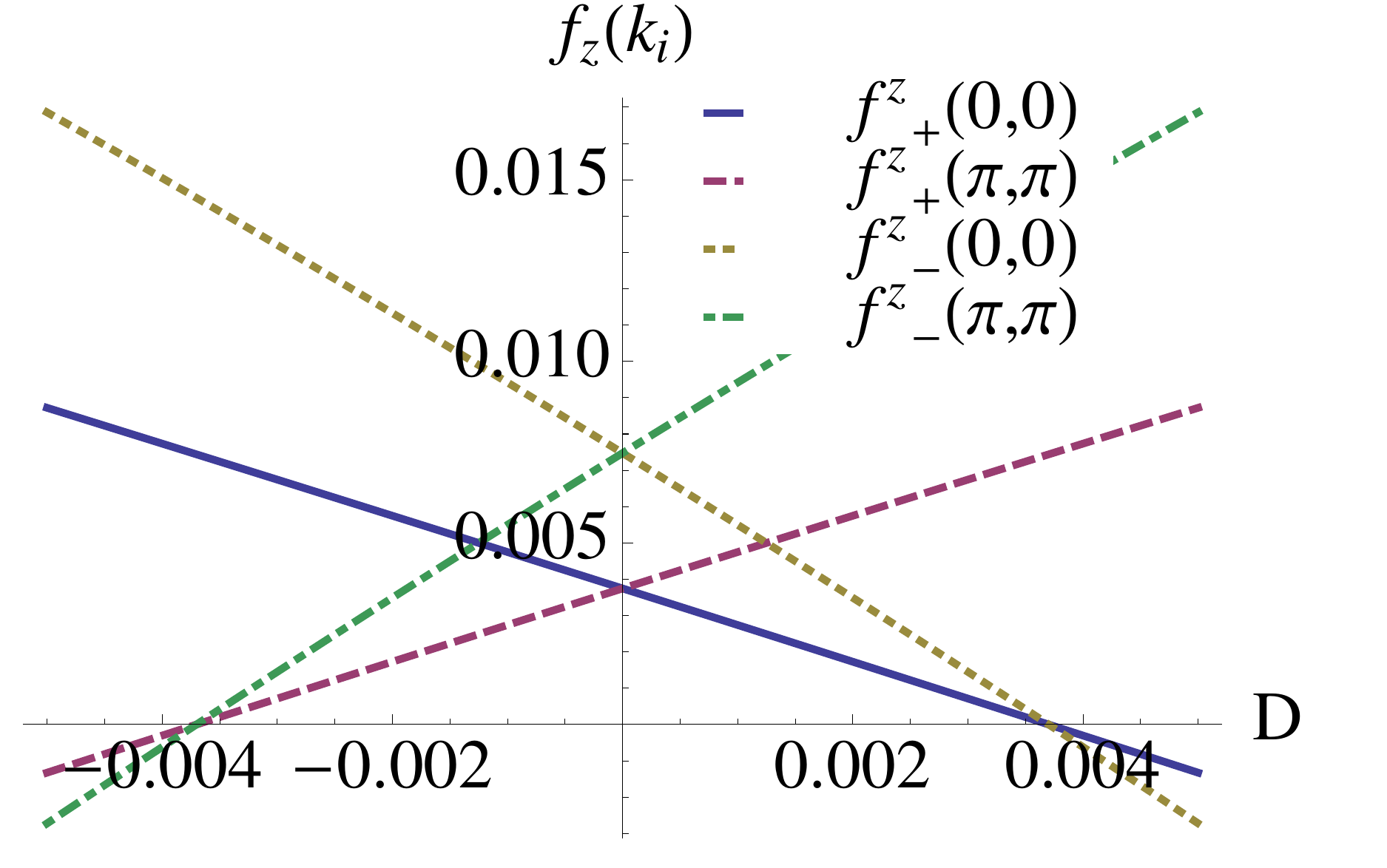}&
\includegraphics[width=.5\linewidth]{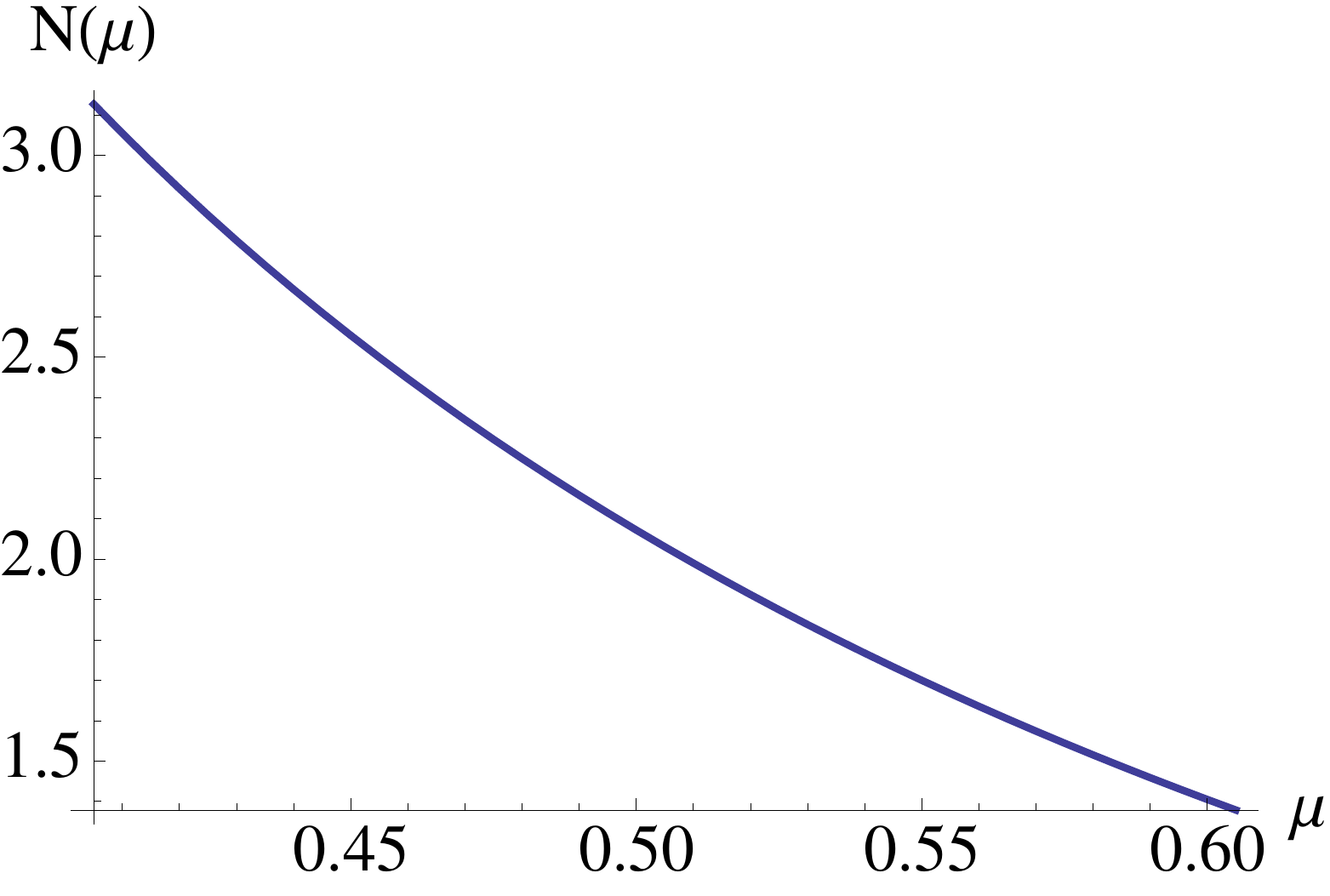}\\
(c)&(d)\\
\end{tabular}
\caption{(Color Online) For $\D/U=.01$: the SO gaps $f^z_\al(\mbk_i)$ as function of $D$ for (a) $\al=-$ and (b) $\al=+$. (c)  Zooming in to the region where the topological transition occurs, at $|D|<.005$, the SO gaps at $(0,0)$ and $(\pi,\pi)$.  (d) The ground state particle per site $N(\mu)$ for the $(1,1)$ filling factor plotted as a function of $\mu$.
}
\label{chern}
\end{center}
\end{figure}

 \section{Conclusion and outlook}

In this work, we have shown that the many body Chern numbers in a generic spin-orbit coupled bosonic Mott insulator can be expressed in terms of the quasiparticle Berry curvature defined in terms of the single particle propagator, which we have computed using a strong-coupling perturbation theory.     This method has several advantages.  It relates the many-body Chern number to quasihole transport properties, somewhat similar to the case of the noninteracting fermions.   It does not require calculating the many body wave function, as in the standard method in the quantum Hall literature, where one imposes twisted periodic boundary conditions on the many body wave function, define a Berry gauge field as a function of the phase changes at the boundary, and compute the Berry flux through the torus of phases.\cite{niuPRB85}  Furthermore, instead of perturbation theory, one can apply various numerical techniques available for calculating single particle Green functions,\cite{yoshidaPRB12}which can be used to check our claim that the leading order of our perturbation theory suffices to determine the Chern numbers.

The topologically nontrivial states that we find are characterized by a spin Chern number, which we expect to be related to the spin Hall conductance.  However, the computation of the spin Hall conductivity when spin is not conserved is a delicate issue, which requires further study.  It is straightforward in the formalism of this paper to include more bands in the hopping Hamiltonian $\hath(\mbk)$, or its time reversed copy.  More internal degrees of freedom may introduce more symmetries and associated topological invariants.  

Besides the Chern numbers, a central result of this work is the calculation of quasiparticle Berry curvature in the presence of strong interactions.  Specifically, it would be interesting to experimentally measure the electric field contribution which is purely interaction generated.   Since it appears in the equation of motion proportional to the band velocity $\mbv_\ksn$, it should be distinguishable from the magnetic component.   Furthermore, in the Fermi liquid, the quasiparticle weight is renormalized by a factor $\sim-\bm{\mcal{E}}\cdot\del\phi$,\cite{shindouPRL06} and we expect this to be valid in the Mott insulator as well.

A realistic comparison with experiments will require a precise computation of the phase diagram for the ground state magnetization, which can be textured.\cite{colePRL12}    As shown in appendix \ref{S1} magnetization textures can readily be included in our theory, and may introduce regions of space that have different SO texture, Berry curvature, and Chern numbers.

To conclude, this work reveals interesting ground state topology and quasiparticle transport properties in bosonic Mott insulators with two dimensional spin-orbit couplings, and we hope this work provides further motivation to study these systems both theoretically and experimentally in cold atom and solid state systems.
 \begin{figure}[b]   
 \centering
   \includegraphics[width=\linewidth]{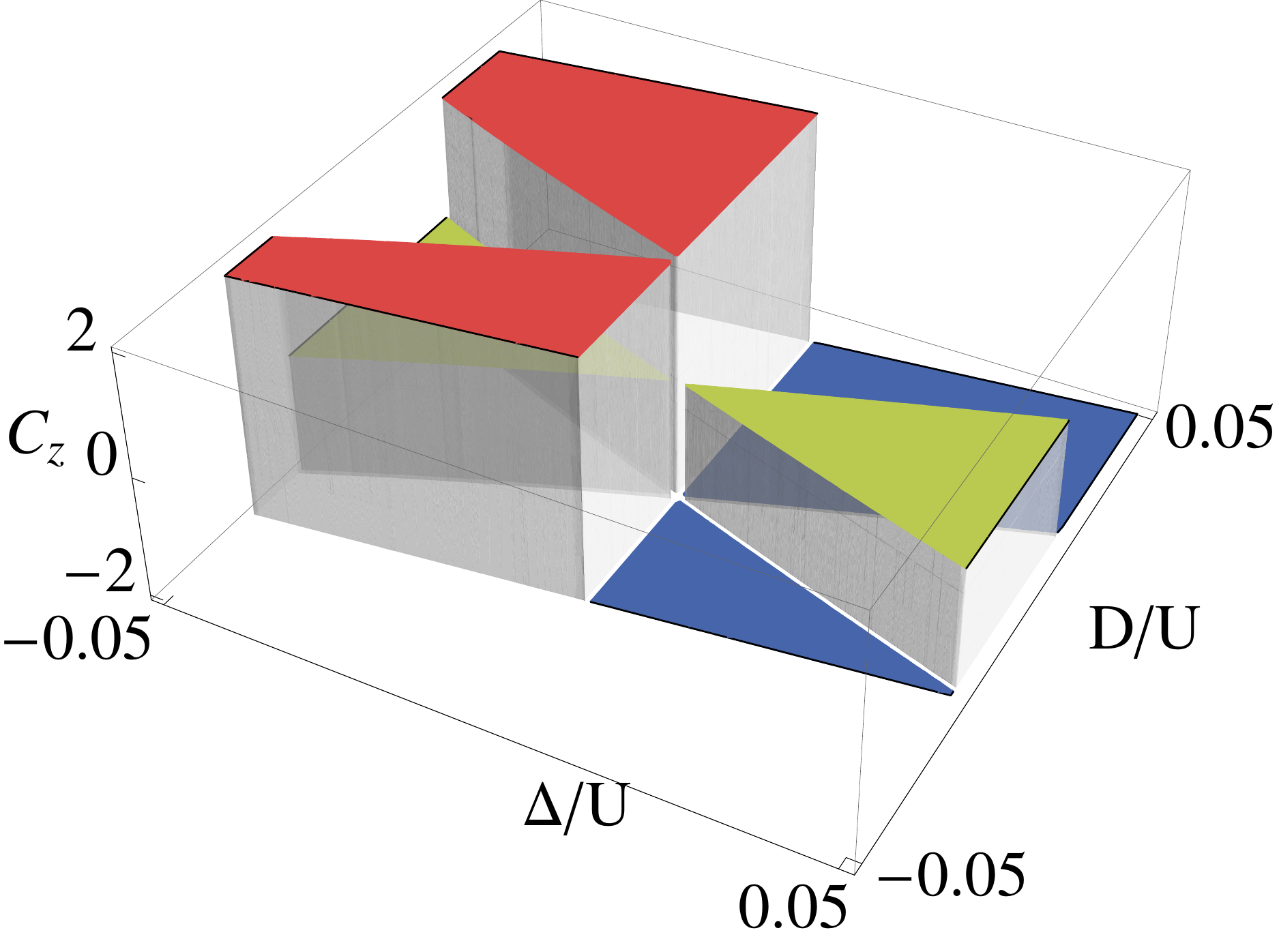} 
   \caption{Phase diagram defined by the spin Chern number $C_z$ as function of the spin independent and spin dependent hopping parameters, $D/U,\D/U$, respectively.  The colors red, green, blue indicate the three values $2,0,-2$, respectively}
   \label{Cz12}
\end{figure}

We thank Henk Stoof, Allan MacDonald, Yaroslav Tserkovnyak, and Ari Turner for stimulating discussions.  This work was supported by Stichting voor Fundamenteel Onderzoek der Materie (FOM), the Netherlands Organization for Scientific
Research (NWO), by the European Research Council (ERC) under the Seventh Framework Program (FP7).

\appendix
\begin{widetext}
\section{Computation of $S_1^{(2)}[\psi]$}
\label{S1}
In this section, we determine the superfluid propagator $G^{(\psi)}$ by computing quadratic part of the effective action $W^{(2)}[\psi]$ [cf.~Eq.~\eqref{W2}]  by expanding $S_1[\vpsi]$, defined in Eq.~\eqref{S1eq}, to second order
\[
S^{(2)}_1[\psi]=
-\ln\int\mcal{D}\va e^{-S_0[\va]}\left[1+\left(\int d\tau \sum_{i,j} t_{ij}(a_i^\dag\psi_{j}+\psi_i^\dag a_{j})\right) +{1\over2}\left(\int d\tau \sum_{i,j} t_{ij}(a_i^\dag\psi_{j}+\psi_i^\dag a_{j})\right)^2+\ldots  \right]
\]
Noting that in the Mott insulator $\<a_\al\>_0=0,\<a_\al^\dag\>_0=0$, $\<a_\al a_\be\>_0=0, \<a_\al^\dag a_\be^\dag\>_0=0$,
where the brackets denote time-ordered expectation value with the unperturbed $H_0$,  dropping $\ln Z_0$  where $Z_0$ is the partition function corresponding to $S_0$, which is independent of $\psi$, and expanding the logarithm, we find 
\begin{align}
S_1^{(2)}[\psi]&=-{1\over2}\left\langle T\left(\int d\tau \sum_{i,j} t_{ij}(a_i^\dag\psi_{j}+\psi_i^\dag a_{j})\right)\left(\int d\tau' \sum_{i',j'} t_{i'j'}(a_{i'}^\dag\psi_{j'}+\psi_{i'}^\dag a_{j'})\right) \right\rangle_0\nn
&=-\int\,d\tau d\tau'\sum_{i,j;i',j'}\psi^\dag_{i'}(\tau')t_{i'j'}\<a_{j'}(\tau')a^\dag_i(\tau)\>_0t_{ij}\psi_j(\tau)\nn
&=\int {d\w\over2\pi}\int{ d\mbk d\mbq\over(2\pi)^4}{\vpsi}^*_{\mbk+\mbq, n}\hat{h}(\mbk+\mbq) \hat{g} (i\omega,\mbq)\hat{h}(\mbk){\vpsi}_{\mbk n}
\label{F2}
\end{align}
\end{widetext}
where $\hat{t}_{ij}=\int\dkk \hath_\mbk e^{i\mbk\cdot(\mbr_i-\mbr_j)}
$,  $\{\mbr_i\}$ are the lattice sites and the unperturbed onsite propagator is given by
\[
 \hat{g}(\mbq)=-\sum_ie^{-i\mbq\cdot\mbr_i}\<{T}a_{i\al}(\tau')a^\dag_{i\beta}(\tau)\>_0
\]
where we allow the possibility of spatial dependence which occurs when the ground state has inhomogeneous magnetization that has a spatially varying angle relative to the fixed quantization axis defined by $a_{i\al}$. The quadratic part of the effective action is thus given by
\begin{align}
\W^{(2)}[\vpsi]&=\int{d\w\over2\pi}\int {d\mbk d\mbq\over(2\pi)^4}\vpsi^\dag(\w,\mbk+\mbq)[-\de(\mbq)\hat{h}(\mbk)\nn
&+\hat{h}(\mbk+\mbq)\hat{g}(i\omega,\mbq)\hat{h}(\mbk)]{\vpsi}(\w,\mbk)
\end{align}

The higher order terms can be computed similarly, however, the evaluation of time-ordered onsite correlation functions with four or more operators becomes a tedious task because as $S_0[\va]$ is quartic in $\va$,  one cannot directly express higher point correlation function in terms of single particle propagators.

\section{Edge states} 
\label{edge}
The inverse Green function provides an effective Schrodinger equation for the quasiparticle, given by the lattice eigenvalue equation
%are a set of complete and orthonormal eigenfunctions.  Because the propagator satisfy  $\sum_j\,\hat{G}^{-1}_{ij}(\w)\hat{G}_{jk}(\w)=\de_{ik}$, mathematically,  $\hatG$ has the eigenfunction expansion\cite {jacksonEM}
%\[ 
%\hat{G}_{ij}(\w)=\sum_n\frac{\vpsi^{(n)}_i\vpsi^{(n)\dag}_j}{\w-\e_n}\,,\quad 
% \]
% satisfying 
 \[
\hatG^{-1}_{ij}(\w)\vpsi_j=\sum_{j}\left[ \w\de_{ij}- \hat{h}_{ij}-\hat{\Sigma}_{ij}(\w) \right]\vec{\psi}_j=0\,,
\]
but, to the leading order in our perturbation theory, the self energy is onsite, so that 
\ben
\sum_{j}\left[\de_{ij}g^{-1}_0(\w)\hat{1}+g^{-1}_z(\w)\hat{\s}_z- \hat{h}_{ij}\right]\vec{\psi}_j=0\,.
\label{qpH}
\een
Together with a set of boundary conditions, one can find a set of wave functions $\vpsi^{(n)}_i$ with eigenvalues $\e_n$.    In addition to the bulk, Bloch states computed above, one can find edges states, for example, in a strip with periodic boundary conditions in $y$ and open boundaries in $x$.

Consider first the case with $g_z^{-1}=0$. Then the eigenfunctions are the same as the wavefunctions of the hopping Hamiltonian, $\hat{h}_{ij}{\vpsi}_j=\e_n^{(1)}{\vpsi}_j$, and near band crossing points $\hath_{ij}$ has the form of a Dirac Hamiltonian with well known edge states.  However, the eigenvalues determined by solving $g^{-1}_0(\w)+\e_n^{(1)}=0$, i.e., inverting the onsite propagator, will produce a particle and hole copies of the edge states spectrum which lie between the bulk particle and hole spin bands.

When $\hat{m}(\w)=g^{-1}_z(\w)\hat{\s}_z$ is nonzero, it provides a ``mass" term, which can cause a  transition to a topologically nontrivial state, as we have seen in the previous, bulk analysis.  However, this term has $\w$ dependence, which has to be solved by satisfying $g^{-1}_0(\w)+\til{\e}_n(\w)=0$, where now $\til{\e}_n^{(1)}(\w)$ are the eigenvalues of $\hath_{ij}+\hat{m}(\w)$.

In practice, the confining potential is smooth, and in a local density approximation, it acts as a local chemical potential which puts the edge of the Mott insulator into a superfluid phase.  However, the computation of the Green function for the superfluid using a strong coupling theory is a delicate matter\cite{senguptaPRA05} which is beyond the scope of this work.

%\bibliography{physics}

\begin{thebibliography}{39}
\expandafter\ifx\csname natexlab\endcsname\relax\def\natexlab#1{#1}\fi
\expandafter\ifx\csname bibnamefont\endcsname\relax
  \def\bibnamefont#1{#1}\fi
\expandafter\ifx\csname bibfnamefont\endcsname\relax
  \def\bibfnamefont#1{#1}\fi
\expandafter\ifx\csname citenamefont\endcsname\relax
  \def\citenamefont#1{#1}\fi
\expandafter\ifx\csname url\endcsname\relax
  \def\url#1{\texttt{#1}}\fi
\expandafter\ifx\csname urlprefix\endcsname\relax\def\urlprefix{URL }\fi
\providecommand{\bibinfo}[2]{#2}
\providecommand{\eprint}[2][]{\url{#2}}

\bibitem[{\citenamefont{Hasan and Kane}(2010)}]{hasanRMP10}
\bibinfo{author}{\bibfnamefont{M.~Z.} \bibnamefont{Hasan}} \bibnamefont{and}
  \bibinfo{author}{\bibfnamefont{C.~L.} \bibnamefont{Kane}},
  \bibinfo{journal}{Rev. Mod. Phys.} \textbf{\bibinfo{volume}{82}},
  \bibinfo{pages}{3045} (\bibinfo{year}{2010});
  \bibinfo{author}{\bibfnamefont{X.-L.} \bibnamefont{Qi}} \bibnamefont{and}
  \bibinfo{author}{\bibfnamefont{S.-C.} \bibnamefont{Zhang}},
  \bibinfo{journal}{Rev. Mod. Phys.} \textbf{\bibinfo{volume}{83}},
  \bibinfo{pages}{1057} (\bibinfo{year}{2011}).

\bibitem[{\citenamefont{Liu et~al.}(2010)\citenamefont{Liu, Liu, Wu, and
  Sinova}}]{liuPRA10}
\bibinfo{author}{\bibfnamefont{X.-J.} \bibnamefont{Liu}},
  \bibinfo{author}{\bibfnamefont{X.}~\bibnamefont{Liu}},
  \bibinfo{author}{\bibfnamefont{C.}~\bibnamefont{Wu}}, \bibnamefont{and}
  \bibinfo{author}{\bibfnamefont{J.}~\bibnamefont{Sinova}},
  \bibinfo{journal}{Phys. Rev. A} \textbf{\bibinfo{volume}{81}},
  \bibinfo{pages}{033622} (\bibinfo{year}{2010}).
  
  \bibitem[{\citenamefont{{Liu} et~al.}(2013{\natexlab{a}})\citenamefont{{Liu},
  {Law}, and {Ng}}}]{liuCM13}
\bibinfo{author}{\bibfnamefont{X.-J.} \bibnamefont{{Liu}}},
  \bibinfo{author}{\bibfnamefont{K.~T.} \bibnamefont{{Law}}}, \bibnamefont{and}
  \bibinfo{author}{\bibfnamefont{T.~K.} \bibnamefont{{Ng}}},
  \bibinfo{journal}{ArXiv:}  \eprint{1304.0291}  (\bibinfo{year}{2013}{\natexlab{a}});
  \bibinfo{author}{\bibfnamefont{X.-J.} \bibnamefont{{Liu}}},
  \bibinfo{author}{\bibfnamefont{K.~T.} \bibnamefont{{Law}}},
  \bibinfo{author}{\bibfnamefont{T.~K.} \bibnamefont{{Ng}}}, \bibnamefont{and}
  \bibinfo{author}{\bibfnamefont{P.~A.} \bibnamefont{{Lee}}},
  \bibinfo{journal}{ArXiv}:   \eprint{1306.5223} (\bibinfo{year}{2013}{\natexlab{b}})
 .

\bibitem[{\citenamefont{Goldman et~al.}(2010)\citenamefont{Goldman, Satija,
  Nikolic, Bermudez, Martin-Delgado, Lewenstein, and Spielman}}]{goldmanPRL10}
\bibinfo{author}{\bibfnamefont{N.}~\bibnamefont{Goldman}},
  \bibinfo{author}{\bibfnamefont{I.}~\bibnamefont{Satija}},
  \bibinfo{author}{\bibfnamefont{P.}~\bibnamefont{Nikolic}},
  \bibinfo{author}{\bibfnamefont{A.}~\bibnamefont{Bermudez}},
  \bibinfo{author}{\bibfnamefont{M.~A.} \bibnamefont{Martin-Delgado}},
  \bibinfo{author}{\bibfnamefont{M.}~\bibnamefont{Lewenstein}},
  \bibnamefont{and} \bibinfo{author}{\bibfnamefont{I.~B.}
  \bibnamefont{Spielman}}, \bibinfo{journal}{Phys. Rev. Lett.}
  \textbf{\bibinfo{volume}{105}}, \bibinfo{pages}{255302}
  (\bibinfo{year}{2010}).

\bibitem{sauPRB11}
J.~D. Sau et. al.,
 { Phys. Rev. B} \textbf{83}, 140510 (2011);
 L.~W. Cheuk et. al., Phys. Rev Lett.
 \textbf{109}, 095302 (2012).

\bibitem[{\citenamefont{Wang et~al.}(2010)\citenamefont{Wang, Qi, and
  Zhang}}]{wangPRL10to}
\bibinfo{author}{\bibfnamefont{Z.}~\bibnamefont{Wang}},
  \bibinfo{author}{\bibfnamefont{X.-L.} \bibnamefont{Qi}}, \bibnamefont{and}
  \bibinfo{author}{\bibfnamefont{S.-C.} \bibnamefont{Zhang}},
  \bibinfo{journal}{Phys. Rev. Lett.} \textbf{\bibinfo{volume}{105}},
  \bibinfo{pages}{256803} (\bibinfo{year}{2010})\bibinfo{author};
  \bibinfo{author}{\bibfnamefont{A.}~\bibnamefont{Shitade}}, \textit{et. al.},
\textit{ibid.}, \textbf{\bibinfo{volume}{102}}, \bibinfo{pages}{256403} (\bibinfo{year}{2009});
D.~Pesin and L.~Balents,
 { Nat. Phys.}, \textbf{6}, 376 (2010);
{\bibfnamefont{S.}~\bibnamefont{Rachel}} \bibnamefont{and}
  \bibinfo{author}{\bibfnamefont{K.}~\bibnamefont{Le~Hur}},
  \bibinfo{journal}{Phys. Rev. B} \textbf{\bibinfo{volume}{82}},
  \bibinfo{pages}{075106} (\bibinfo{year}{2010});
\bibinfo{author}{\bibfnamefont{E.~M.} \bibnamefont{Stoudenmire}},
\textit{et. al.},  \textit{ibid.},   \textbf{\bibinfo{volume}{84}}, \bibinfo{pages}{014503}
  (\bibinfo{year}{2011});
\bibinfo{author}{\bibfnamefont{Y.}~\bibnamefont{Tada}} \textit{et. al.},
\textit{ibid.},  \textbf{\bibinfo{volume}{85}},
  \bibinfo{pages}{165138} (\bibinfo{year}{2012}).
  
 \bibitem{yoshidaPRB12} 
\bibinfo{author}{\bibfnamefont{T.}~\bibnamefont{Yoshida}},
  \bibinfo{author}{\bibfnamefont{S.}~\bibnamefont{Fujimoto}}, \bibnamefont{and}
  \bibinfo{author}{\bibfnamefont{N.}~\bibnamefont{Kawakami}},
 \bibinfo{journal}{Phys. Rev. B} \textbf{\bibinfo{volume}{85}},
  \bibinfo{pages}{125113} (\bibinfo{year}{2012});
\bibinfo{author}{\bibfnamefont{T.}~\bibnamefont{Yoshida}} \textit{et. al.},
\textit{ibid.},  \textbf{\bibinfo{volume}{87}},
  \bibinfo{pages}{085134} (\bibinfo{year}{2013});
  \bibinfo{author}{\bibfnamefont{S.}~\bibnamefont{Miyakoshi}} \bibnamefont{and}
  \bibinfo{author}{\bibfnamefont{Y.}~\bibnamefont{Ohta}},
 \textit{ibid.},  \textbf{\bibinfo{volume}{87}},
  \bibinfo{pages}{195133} (\bibinfo{year}{2013}).


\bibitem[{\citenamefont{Chen et~al.}(2011)\citenamefont{Chen, Liu, and
  Wen}}]{chenPRB11}
\bibinfo{author}{\bibfnamefont{X.}~\bibnamefont{Chen}},
  \bibinfo{author}{\bibfnamefont{Z.-C.} \bibnamefont{Gu}},
  \bibinfo{author}{\bibfnamefont{Z.-X.} \bibnamefont{Liu}}, \bibnamefont{and}
  \bibinfo{author}{\bibfnamefont{X.-G.} \bibnamefont{Wen}},
  \bibinfo{journal}{Science} \textbf{\bibinfo{volume}{338}},
  \bibinfo{pages}{1604} (\bibinfo{year}{2012});
  \bibinfo{author}{\bibfnamefont{X.}~\bibnamefont{Chen}},
  \bibinfo{author}{\bibfnamefont{Z.-X.} \bibnamefont{Liu}}, \bibnamefont{and}
  \bibinfo{author}{\bibfnamefont{X.-G.} \bibnamefont{Wen}},
  \bibinfo{journal}{Phys. Rev. B} \textbf{\bibinfo{volume}{84}},
  \bibinfo{pages}{235141} (\bibinfo{year}{2011}); 
\bibinfo{author}{\bibfnamefont{T.}~\bibnamefont{Grover}} \bibnamefont{and}
  \bibinfo{author}{\bibfnamefont{A.}~\bibnamefont{Vishwanath}},
\textit{ibid.},   \textbf{\bibinfo{volume}{87}},
  \bibinfo{pages}{045129} (\bibinfo{year}{2013});
\bibinfo{author}{\bibfnamefont{M.}~\bibnamefont{Levin}} \bibnamefont{and}
  \bibinfo{author}{\bibfnamefont{Z.-C.} \bibnamefont{Gu}},
\textit{ibid.},    \textbf{\bibinfo{volume}{86}},
  \bibinfo{pages}{115109} (\bibinfo{year}{2012});
\bibinfo{author}{\bibfnamefont{Y.-M.} \bibnamefont{Lu}} \bibnamefont{and}
  \bibinfo{author}{\bibfnamefont{A.}~\bibnamefont{Vishwanath}},
\textit{ibid.}, \textbf{\bibinfo{volume}{86}},
  \bibinfo{pages}{125119} (\bibinfo{year}{2012});
\bibinfo{author}{\bibfnamefont{M.~A.} \bibnamefont{Metlitski}},
  \bibinfo{author}{\bibfnamefont{C.~L.} \bibnamefont{Kane}}, \bibnamefont{and}
  \bibinfo{author}{\bibfnamefont{M.~P.~A.} \bibnamefont{Fisher}}, ArXiv:\eprint{1302.6535} (\bibinfo{year}{2013});
\bibinfo{author}{\bibfnamefont{C.}~\bibnamefont{Wang}} \bibnamefont{and}
  \bibinfo{author}{\bibfnamefont{T.}~\bibnamefont{Senthil}}, ArXiv: \eprint{1302.6234} (\bibinfo{year}{2013}),
\bibinfo{author}{\bibfnamefont{C.}~\bibnamefont{Xu}} \bibnamefont{and}
  \bibinfo{author}{\bibfnamefont{T.}~\bibnamefont{Senthil}}, ArXiv: \eprint{1301.6172} (\bibinfo{year}{2013}).
 
\bibitem{senthilPRL13}
\bibinfo{author}{\bibfnamefont{T.}~\bibnamefont{Senthil}} \bibnamefont{and}
  \bibinfo{author}{\bibfnamefont{M.}~\bibnamefont{Levin}},
  \bibinfo{journal}{Phys. Rev. Lett.} \textbf{\bibinfo{volume}{110}},
  \bibinfo{pages}{046801} (\bibinfo{year}{2013}).
  
  
\bibitem[{\citenamefont{Greiner et~al.}(2002)\citenamefont{Greiner, Mandel,
  Esslinger, Hansch, and Bloch}}]{greinerNAT02}
\bibinfo{author}{\bibfnamefont{M.}~\bibnamefont{Greiner}} \textit{et. al.},
  \bibinfo{journal}{Nature} \textbf{\bibinfo{volume}{415}}, \bibinfo{pages}{39}
  (\bibinfo{year}{2002}).

\bibitem[{\citenamefont{Lin et~al.}(2011)\citenamefont{Lin, Jimenez-Garcia, and
  Spielman}}]{linNAT11}
\bibinfo{author}{\bibfnamefont{Y.~J.} \bibnamefont{Lin}},
  \bibinfo{author}{\bibfnamefont{K.}~\bibnamefont{Jimenez-Garcia}},
  \bibnamefont{and} \bibinfo{author}{\bibfnamefont{I.~B.}
  \bibnamefont{Spielman}}, \bibinfo{journal}{Nature}
  \textbf{\bibinfo{volume}{471}}, \bibinfo{pages}{83} (\bibinfo{year}{2011}).



\bibitem[{\citenamefont{Wong and Duine}(2013)}]{wongPRL13}
\bibinfo{author}{\bibfnamefont{C.~H.} \bibnamefont{Wong}} \bibnamefont{and}
  \bibinfo{author}{\bibfnamefont{R.~A.} \bibnamefont{Duine}},
  \bibinfo{journal}{Phys. Rev. Lett.} \textbf{\bibinfo{volume}{110}},
  \bibinfo{pages}{115301} (\bibinfo{year}{2013}).


\bibitem[{\citenamefont{Price and Cooper}(2012)}]{pricePRA12}
\bibinfo{author}{\bibfnamefont{H.~M.} \bibnamefont{Price}} \bibnamefont{and}
  \bibinfo{author}{\bibfnamefont{N.~R.} \bibnamefont{Cooper}},
  \bibinfo{journal}{Phys. Rev. A} \textbf{\bibinfo{volume}{85}},
  \bibinfo{pages}{033620} (\bibinfo{year}{2012}).

\bibitem[{\citenamefont{Bernevig et~al.}(2006)\citenamefont{Bernevig, Hughes,
  and Zhang}}]{bernevigSCI06}
\bibinfo{author}{\bibfnamefont{B.~A.} \bibnamefont{Bernevig}},
  \bibinfo{author}{\bibfnamefont{T.~L.} \bibnamefont{Hughes}},
  \bibnamefont{and} \bibinfo{author}{\bibfnamefont{S.-C.} \bibnamefont{Zhang}},
  \bibinfo{journal}{Science} \textbf{\bibinfo{volume}{314}},
  \bibinfo{pages}{1757} (\bibinfo{year}{2006}).

\bibitem{TretiakovPRB07}
O.~A. Tretiakov and O.~Tchernyshyov.
\newblock {\em Phys. Rev. B } \textbf{75} 012408, (2007).



\bibitem[{\citenamefont{Raghu et~al.}(2008)\citenamefont{Raghu, Qi, Honerkamp,
  and Zhang}}]{raghuPRL08}
\bibinfo{author}{\bibfnamefont{S.}~\bibnamefont{Raghu}},
  \bibinfo{author}{\bibfnamefont{X.-L.} \bibnamefont{Qi}},
  \bibinfo{author}{\bibfnamefont{C.}~\bibnamefont{Honerkamp}},
  \bibnamefont{and} \bibinfo{author}{\bibfnamefont{S.-C.} \bibnamefont{Zhang}},
  \bibinfo{journal}{Phys. Rev. Lett.} \textbf{\bibinfo{volume}{100}},
  \bibinfo{pages}{156401} (\bibinfo{year}{2008}).

\bibitem[{\citenamefont{Ishikawa and Matsuyama}(1987)}]{ishikawaNPB87}
\bibinfo{author}{\bibfnamefont{K.}~\bibnamefont{Ishikawa}} \bibnamefont{and}
  \bibinfo{author}{\bibfnamefont{T.}~\bibnamefont{Matsuyama}},
  \bibinfo{journal}{Nuclear Physics B} \textbf{\bibinfo{volume}{280}},
  \bibinfo{pages}{523 } (\bibinfo{year}{1987});
  \bibitem{HaldanePRL04}
F.~D.~M. Haldane, 
\newblock {\em Phys. Rev. Lett.}\textbf{93} 206602 (2004).



\bibitem[{\citenamefont{Shindou and Balents}(2006)}]{shindouPRL06}
\bibinfo{author}{\bibfnamefont{R.}~\bibnamefont{Shindou}} \bibnamefont{and}
  \bibinfo{author}{\bibfnamefont{L.}~\bibnamefont{Balents}},
  \bibinfo{journal}{Phys. Rev. Lett.} \textbf{\bibinfo{volume}{97}},
  \bibinfo{eid}{216601}  (\bibinfo{year}{2006});
 { Phys. Rev. B}, \textbf{77}, 035110 (2008).

\bibitem[{\citenamefont{Zhu et~al.}(2006)\citenamefont{Zhu, Fu, Wu, Zhang, and
  Duan}}]{zhuPRL06}
\bibinfo{author}{\bibfnamefont{S.-L.} \bibnamefont{Zhu}},
  \bibinfo{author}{\bibfnamefont{H.}~\bibnamefont{Fu}},
  \bibinfo{author}{\bibfnamefont{C.-J.} \bibnamefont{Wu}},
  \bibinfo{author}{\bibfnamefont{S.-C.} \bibnamefont{Zhang}}, \bibnamefont{and}
  \bibinfo{author}{\bibfnamefont{L.-M.} \bibnamefont{Duan}},
  \bibinfo{journal}{Phys. Rev. Lett.} \textbf{\bibinfo{volume}{97}},
  \bibinfo{pages}{240401} (\bibinfo{year}{2006}).

\bibitem[{\citenamefont{Matsumoto and Murakami}(2011)}]{matsumotoPRL11}
\bibinfo{author}{\bibfnamefont{R.}~\bibnamefont{Matsumoto}} \bibnamefont{and}
  \bibinfo{author}{\bibfnamefont{S.}~\bibnamefont{Murakami}},
  \bibinfo{journal}{Phys. Rev. Lett.} \textbf{\bibinfo{volume}{106}},
  \bibinfo{pages}{197202} (\bibinfo{year}{2011}).

\bibitem[{\citenamefont{Wong and Tserkovnyak}(2011)}]{wongPRB11}
\bibinfo{author}{\bibfnamefont{C.~H.} \bibnamefont{Wong}} \bibnamefont{and}
  \bibinfo{author}{\bibfnamefont{Y.}~\bibnamefont{Tserkovnyak}},
  \bibinfo{journal}{Phys. Rev. B} \textbf{\bibinfo{volume}{84}},
  \bibinfo{pages}{115209} (\bibinfo{year}{2011}).

\bibitem[{\citenamefont{Fisher et~al.}(1989)\citenamefont{Fisher, Weichman,
  Grinstein, and Fisher}}]{fisherPRB89}
\bibinfo{author}{\bibfnamefont{M.~P.~A.} \bibnamefont{Fisher}} \textit{et. al.},
 \bibinfo{journal}{Phys. Rev. B}
  \textbf{\bibinfo{volume}{40}}, \bibinfo{pages}{546} (\bibinfo{year}{1989}).

\bibitem[{\citenamefont{van Oosten et~al.}(2001)\citenamefont{van Oosten,
  van~der Straten, and Stoof}}]{oostenPRA01}
\bibinfo{author}{\bibfnamefont{D.}~\bibnamefont{van Oosten}},
  \bibinfo{author}{\bibfnamefont{P.}~\bibnamefont{van~der Straten}},
  \bibnamefont{and} \bibinfo{author}{\bibfnamefont{H.~T.~C.}
  \bibnamefont{Stoof}}, \bibinfo{journal}{Phys. Rev. A}
  \textbf{\bibinfo{volume}{63}}, \bibinfo{pages}{053601}
  (\bibinfo{year}{2001}).
  
  

\bibitem[{\citenamefont{Isacsson et~al.}(2005)\citenamefont{Isacsson, Cha,
  Sengupta, and Girvin}}]{isacssonPRB05}
\bibinfo{author}{\bibfnamefont{A.}~\bibnamefont{Isacsson}},
  \bibinfo{author}{\bibfnamefont{M.-C.} \bibnamefont{Cha}},
  \bibinfo{author}{\bibfnamefont{K.}~\bibnamefont{Sengupta}}, \bibnamefont{and}
  \bibinfo{author}{\bibfnamefont{S.~M.} \bibnamefont{Girvin}},
  \bibinfo{journal}{Phys. Rev. B} \textbf{\bibinfo{volume}{72}},
  \bibinfo{pages}{184507} (\bibinfo{year}{2005}).

\bibitem[{\citenamefont{Niu et~al.}(1985)\citenamefont{Niu, Thouless, and
  Wu}}]{niuPRB85}
\bibinfo{author}{\bibfnamefont{Q.}~\bibnamefont{Niu}},
  \bibinfo{author}{\bibfnamefont{D.~J.} \bibnamefont{Thouless}},
  \bibnamefont{and} \bibinfo{author}{\bibfnamefont{Y.-S.} \bibnamefont{Wu}},
  \bibinfo{journal}{Phys. Rev. B} \textbf{\bibinfo{volume}{31}},
  \bibinfo{pages}{3372} (\bibinfo{year}{1985});
\bibinfo{author}{\bibfnamefont{X.-L.} \bibnamefont{Qi}},
  \bibinfo{author}{\bibfnamefont{Y.-S.} \bibnamefont{Wu}}, \bibnamefont{and}
  \bibinfo{author}{\bibfnamefont{S.-C.} \bibnamefont{Zhang}},
  \textit{ibid.}, \textbf{\bibinfo{volume}{74}},
  \bibinfo{pages}{045125} (\bibinfo{year}{2006}).

\bibitem[{\citenamefont{Cole et~al.}(2012)\citenamefont{Cole, Zhang,
  Paramekanti, and Trivedi}}]{colePRL12}
\bibinfo{author}{\bibfnamefont{W.~S.} \bibnamefont{Cole}} et. al.,
  \bibinfo{journal}{Phys. Rev. Lett.} \textbf{\bibinfo{volume}{109}},
  \bibinfo{pages}{085302} (\bibinfo{year}{2012});
J.~Radi\ifmmode~\acute{c}\else \'{c}\fi{} et. al.,
\textit{ibid}. \textbf{09}, 085303 (2012);
L.~He et. al., {Phys. Rev. A} \textbf{86} 043620 (2012);
Z.~Cai et. al.,  { Phys. Rev. A}, \textbf{85} 061605 (2012).

\bibitem[{\citenamefont{Sengupta and Dupuis}(2005)}]{senguptaPRA05}
\bibinfo{author}{\bibfnamefont{K.}~\bibnamefont{Sengupta}} \bibnamefont{and}
  \bibinfo{author}{\bibfnamefont{N.}~\bibnamefont{Dupuis}},
  \bibinfo{journal}{Phys. Rev. A} \textbf{\bibinfo{volume}{71}},
  \bibinfo{pages}{033629} (\bibinfo{year}{2005}).

\bibitem[{\citenamefont{Nielsen and Ninomiya}(1981)}]{nielsenNPB81b}
\bibinfo{author}{\bibfnamefont{H.~B.} \bibnamefont{Nielsen}} \bibnamefont{and}
  \bibinfo{author}{\bibfnamefont{M.}~\bibnamefont{Ninomiya}},
  \bibinfo{journal}{Nucl. Phys. B} \textbf{\bibinfo{volume}{193}},
  \bibinfo{pages}{173 } (\bibinfo{year}{1981}).
  
\bibitem[{\citenamefont{Landau and Lifshitz}(1980)}]{landauSP2}
\bibinfo{author}{\bibfnamefont{L.}~\bibnamefont{Landau}} \bibnamefont{and}
  \bibinfo{author}{\bibfnamefont{E.}~\bibnamefont{Lifshitz}},
  \emph{\bibinfo{title}{Statistical Physics, Part 2}}
  (\bibinfo{publisher}{Pergamon Press}, \bibinfo{year}{1980}).

\bibitem[{\citenamefont{Gra\ss{} et~al.}(2011)\citenamefont{Gra\ss{}, Saha,
  Sengupta, and Lewenstein}}]{grassPRA11}
\bibinfo{author}{\bibfnamefont{T.}~\bibnamefont{Gra\ss{}}},
  \bibinfo{author}{\bibfnamefont{K.}~\bibnamefont{Saha}},
  \bibinfo{author}{\bibfnamefont{K.}~\bibnamefont{Sengupta}}, \bibnamefont{and}
  \bibinfo{author}{\bibfnamefont{M.}~\bibnamefont{Lewenstein}},
  \bibinfo{journal}{Phys. Rev. A} \textbf{\bibinfo{volume}{84}},
  \bibinfo{pages}{053632} (\bibinfo{year}{2011}).

\end{thebibliography}
\end{document}